# An analysis of (the lack of) slip transfer between near-cube oriented grains in pure Al

T. R Bieler[1,2], R. Alizadeh[1] M. Peña-Ortega[3], and J. LLorca[1,3]

[1] IMDEA Materials Institute, c/Eric Kandel 2, 28906 Getafe, Madrid, Spain.
[2] Department of Chemical Engineering and Materials Science, Michigan State University, East Lansing, MI 48824-1226, USA.
[3] Department of Materials Science, Polytechnic University of Madrid. E.T.S. de Ingenieros de Caminos, 28040 Madrid, Spain.

**Abstract**

Slip transfer across grain boundaries was studied in annealed polycrystalline Al foils deformed in uniaxial tension by means of the analysis of the slip traces on the specimen surface. Grain orientations and selected grain boundary misorientations were measured on both surfaces of the sample using electron back-scattered diffraction mapping. It was found that most of the grains were within 15° of a cube orientation and approximately half of the grains percolated through the specimen thickness. The Luster-Morris $m'$ parameter (that can be computed from the surface grain orientation) was used to assess the likelihood of slip transfer across boundaries. It was found that transfer across grain boundaries was rare in near-cube oriented grains, and convincing evidence was only found when $m' > 0.97$, which corresponds to low-angle boundaries with <15º misorientation. This behavior was explained by the presence of many active slip slips in near-cube oriented grains that favor self-accommodation of the grain shape to the evolving boundary conditions imposed by neighboring grains instead of promoting slip transfer across the boundary. These results indicate that the alignment between slip planes and slip directions across the boundary is not the only important metric to determine the threshold for slip transfer, as the particular details of deformation in each grain (such as the number of available slip systems) also must be considered.

# 1. Introduction

The heterogeneous deformation of polycrystals depends upon the local microstructure, where differences in grain orientations lead to different rates of deformation in each grain as constrained by the deformation of their neighboring grains (Delaire et al., 2000, Bieler et al., 2014). Deformation near grain boundaries mediates the strain jump and there is ample experimental evidence showing that grain boundaries may act as strong barriers for dislocations, which are stopped at the boundary leading to the formation of pile-ups, or are easily transferred to the neighbor grain leading to the propagation of deformation on a different slip system (Bieler et al., 2014; Bayerschen et al., 2016; Malyar et al., 2017; Hémery et al., 2018). Moreover, recent strain measurements by digital image correlation have shown that local strains near annealing twin boundaries may exceed several times the average strain (Stinville et al., 2015). Given that grain boundaries are often preferential locations for damage nucleation (Boehlert et al., 2012; Muñoz-Moreno et al., 2013) and that strain energy is often used as a metric to identify potential damage sites (e.g. Sangid et al., 2011; Musinki and McDowell, 2016; Wan et al. 2016; Cruzado et al., 2018; Chen et al, 2018), the need to accurately predict localized deformation at and near grain boundaries seems quite important to understand and predict accurately the mechanical performance of polycrystals.

Computational homogenization, using either mean-field approximations or full-field methods based on the finite element method or the fast Fourier transform, is the current strategy to simulate the mechanical behaviour of polycrystals including a sufficiently large number of grains with realistic shapes, orientation and misorientation distributions (e.g. Lebensohn and Tomé, 1993; Roters et al., 2010; Segurado et al., 2018). The mechanical behaviour of each crystal within the polycrystal follows the crystal plasticity formalism that constrains the process of plastic deformation to occur by crystallographic slip systems, which provides an accurate and physically-based representation of this phenomenon at the microscopic level within each grain. Nevertheless, standard simulations of polycrystals using this approach do not take into account the formation of pile-ups at opaque grain boundaries nor the generation of geometrically necessary dislocations near the grain boundaries to accommodate the inhomogeneous deformation gradients. As a result, comparisons between experimental measurements of the local strain and accurate simulations of the same experiments show partial agreement and rather variable disagreements typically occur near grain boundaries (e.g. Yang et al., 2011; Lim et al., 2015; Guery et al., 2016). Thus, it is not possible to make reliable predictions of damage thresholds without introduction of grain boundary properties into mesoscale models.

Atomistic modelling has provided very useful insights on the dislocation/grain boundary interactions as well as on the mechanisms of slip transfer (e.g. Bayerschen et al., 2016, Spearot and Sangid, 2014; Tschopp et al., 2008; Tsuru et al., 2016; Zhang et al., 2016) but it is not possible to simulate plastic deformation of polycrystals at the atomistic scale due to the intrinsic limitations of this modelling strategy. Rigorous modeling of grain boundaries can be done based upon strictly continuum perspectives (Gurtin 2002; Abu Al-Rub and Voyiadjis, 2006; Gurtin 2008; Van Beers et al., 2013; Bond and Zikry, 2017) but it leads to very complex models that are not suitable to simulate very large models containing many grains. Simpler dislocation-based crystal plasticity models take into account the generation of geometrically necessary dislocations near the grain boundaries to accommodate the anisotropic deformation of the grains (Busso et al., 2000; Ma et al., 2006; Lim et al., 2011). These models can predict size effects associated with grain boundaries but do not include the influence of dislocation pile-ups that are associated with opaque grain boundaries. This latter mechanism is necessary to predict accurately the Hall-Petch effect in FCC polycrystals, as recently shown (Haouala et al., 2018, Rubio et al., 2019, Lim et al., 2011).

In order to introduce the grain boundary properties into mesoscale models (through either phenomenological or physically-based models), it is necessary to understand the deformation processes that actually occur at grain boundaries and, in particular, the factors that control slip transfer across the boundary (Lee et al., 1989; Luster and Morris, 1995; Koning et al., 2002; Dewald and Curtin 2007; Bieler et al., 2014; Kacher et al., 2014; Tsuru et al., 2016; Bayerschen et al., 2016). Slip transfer across grain boundaries enables long-range shear bands to form (Abuzaid et al., 2016), which affect the length of eventual fatigue crack formation, and also can lead to twin formation (Wang et al., 2010; Eftink et al., 2017). Moreover, slip localization parallel to annealing twins can lead to a preferential site for fatigue crack nucleation in Ni-based superalloys (Stinville et al., 2017).

Rules for slip transfer have been identified based upon bicrystal experiments and transmission electron microscopy investigations of polycrystals (Livingston and Chalmers, 1957; Lee et al., 1989; Luster and Morris, 1995; Bayerschen et al., 2016), indicating that transmission is more likely when the slip plane and slip direction are closely aligned, and the angle between the slip planes in the boundary is small. Other important factors are minimizing the residual Burgers vector content left within the boundary following a transmission event, and a sufficiently high resolved stress to drive the transmitted slip. Nevertheless, atomistic and dislocation dynamics simulations have shown that the process of dislocation absorption/emission from/to a grain boundary is very complex and depends on many factors. For instance, dislocation pileups can be prevented at a transparent boundary, yet accumulation of defects within the boundary, such as residual Burgers vectors resulting from slip transmission, may increase the energy of the boundary, making it less amenable to slip transmission (Chandra et al., 2015, 2016) such that slip transmission properties evolve with strain (Xu et al., 2016). Moreover, the significance of slip transmission in the mesoscale deformation process depends on whether activated slip systems are effectively transparent with respect to the slip systems in neighboring grains. Thus, operation of slip transmission depends on both grain and grain boundary misorientation distributions. From the above, there are at least two interdependent thresholds for slip transmission, one that is associated with a stress build up from a dislocation pileup for a given misorientation, and another that depends on magnitude of the slip system misalignment. Also, slip transmission is a subset of slip transfer, where transmission refers to passing a dislocation through a boundary; slip transfer could also occur by absorption of dislocations into a boundary, and emission of dislocations into the neighboring grain, a distinction that is discussed in greater depth in the review by Bayerschen et al. (2016). As this distinction cannot be resolved in mesoscale observations, the term 'slip transfer' is used to be more general in this paper.

While there is much discussion of slip transfer in the literature, there is relatively little experimental quantification of thresholds for observed slip transfer. From the experimental perspective, thresholds for slip transfer have been identified in nano-indentation experiments (Su et al., 2016; Xiao et al., 2017), and thresholds for specific boundaries have been identified in cantilever beam bending tests [Ding et al., 2016]. Analysis of slip stimulated mechanical twinning in Ti using surface electron back-scatter diffraction (EBDS) orientation measurements showed that slip transfer that nucleated mechanical twins was probable when $m' = \cos\psi \cos\kappa > 0.9$ (Wang et al., 2010), where $\psi$ and $\kappa$ are the angles between the slip plane normal and the Burgers vector directions, respectively, for each combination of slip systems[1]. Hémery et al. (2017) carried out a detailed statistical analysis of slip transfer in a Ti-6Al-4V alloy using the same strategy and found that $m'$ was a good indicator of the probability of slip transfer for the different slip systems (basal, pyramidal and prismatic) together with a high resolved shear stress on the outgoing slip system, and slip transfer was observed for $m'$ values greater than 0.6. Nevertheless, similar information is difficult to find for cubic systems and this is the main objective of this investigation. To this end, slip transfer was analysed in a pure Al polycrystal in which most of the grains are within 15° of a cube orientation. Such textures are commonly found in recrystallized face-centered cubic (FCC) metals and alloys, so identification of thresholds for slip transfer in such a microstructure will be valuable to enable simulations that can address how slip transfer affects the kinematics of deformation, and hence better predict the evolution of local strains near grain boundaries.

2. **Experimental procedure and analysis methods**

High purity Al foils (99.9995%) with a thickness of 200 μm were purchased from Alfa Aesar. The initial material was in the annealed condition with average grain size of about 195 ± 30 μm. It was further annealed at 360°C for different annealing times (between 15 and 1440 minutes) to promote grain growth. The grain growth saturated after 60 min of annealing, so all the samples were annealed for 60 min, resulting in an average grain size of 390 ± 30 μm (Fig. 1). It is likely that the rolling direction (which was unknown) corresponds with the longer grain dimension.

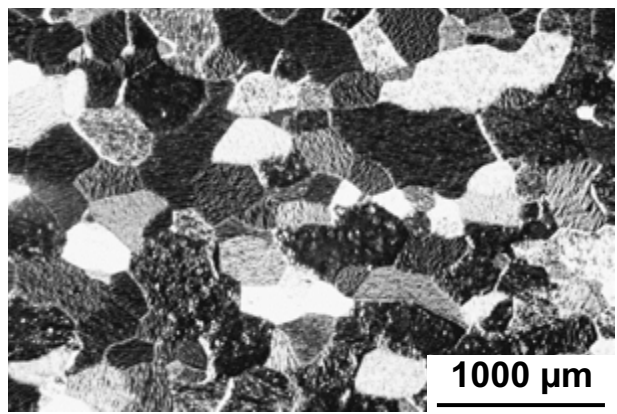

Fig. 1. Microstructure of the Al foil showing the grains after annealing at 360ºC for 60 minutes.

---

[1] This particular geometrical parameter for slip transfer does not require knowledge of the grain boundary inclination, making it practical for surface measurements.

Tensile samples were cut from the foils by electro discharge machining according to the dimensions given in Fig. 2. Because the surface was already very smooth, only electro-polishing was done to obtain a suitable surface for EBSD measurements. Both surfaces of the sample were electro-polished at -10 °C with applied voltage of 30 V using a solution of 10 vol.% perchloric acid with 90 vol.% ethanol for 200 s. They were examined before testing using a field emission gun scanning electron microscope (Helios Nanolab 600i) equipped with an Oxford-HKL electron back-scatter diffraction system using a step size of 5 microns to identify the crystal orientation and microstructure. The samples were tested in tension at room temperature using a micro-tensile testing machine (Kammrath and Weiss) up to an applied strain of 4%. The tensile tests were performed under constant cross head speed of 5.0 μm/s, which corresponds to initial strain rate of $10^{-3}$/s. A fixture was designed and built to ideally enable the analysis of the deformed samples in the scanning electron microscope using secondary and back-scatter electron detectors (without removing the sample from the fixture) to determine the slip traces as well as the changes in the grain orientation. As the specimen was investigated multiple times in the microscope, a brief amount of chemical polishing with dilute Keller's solution for a few seconds was done prior to most electron microscopy sessions to improve the EBSD pattern quality. Unfortunately, this resulted in some etch pit artefacts, which in addition to residual surface contamination and accumulated scratches resulting from handling, features of interest were partially obscured in some areas.

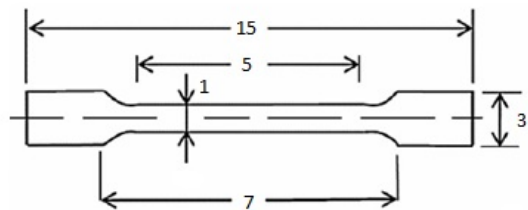

Fig. 2. Dimensions of the tensile specimens machined from the Al foils. All dimensions are in mm.

The EBSD data were converted to a hexagonal mesh format readable by the TSL Analysis software version 7 using a Matlab code. The data were cleaned up using a confidence index criterion (using the inverse of the mean angular deviation (MAD) parameter from the HKL data) to replace pixels with a low confidence index with a neighbor pixel with a higher value in the same grain. Grains smaller than 10 pixels were eliminated using a dilation process, and these processes changed about 10% of the pixels. Finally, the asymmetric domain method was used to have the Euler angles for each grain reside in the same symmetric subset of orientation space. Type 2 grain files and reconstructed boundary files were exported and further analyzed using a Matlab code that computed values for the slip transfer parameter $m' = \cos\psi \cos\kappa$. The values of $\psi$ and $\kappa$ were computed from the scalar product of the plane normal (*hkl*) and slip direction (*uvw*) in neighboring grains expressed in the sample coordinate system using **X** = **xg**, where **X** represents the crystal coordinate system vector **x** in the sample coordinate system, and **g** is the orientation matrix computed from the Bunge Euler angles. The code also generates pairs of unit cells in relative positions defined by the grain boundary trace for slip system combinations with high $m'$ values with the slip systems and slip traces identified visually. Plane traces of the slip planes were computed from the cross product of the slip plane normal **n** (in the sample coordinate direction) with the specimen *Z* direction, i.e. **N** × [001], where **N** = **ng**. The $m'$ values are computed from the scalar products of $\cos\psi = N_iN_j$ and $\cos\kappa = B_iB_j$, where **B** = **bg**, is the slip direction, **b** the Burgers vector and *i* and *j* refer to the grains on either side of the boundary. The tables of $m'$ values (which are described in more detail below) were calculated for all boundaries based upon the average grain orientation. Orientation pairs close to the boundary were extracted in some boundaries to determine if the local crystal orientations differed significantly from the average grain orientations, and if they affected the analysis of $m'$ and the interpretation of slip transfer.

3.  **Results**

Tensile samples with the same orientation with respect to the foil were deformed to approximately 4% strain and then the surface was characterized using scanning electron microscopy (SEM) and EBSD. The analysis described in detail in this paper focused on two regions on sample #1, one on the right side and the other one on the left side of the tensile sample, as illustrated in Figs. 3 and 4, respectively (the middle 100 μm had more handling damage, so no analysis was done there). Analysis was also done on sample #4 in the same manner, and the overall results from this sample are similar and not otherwise described, but are included to provide better statistical sampling. Slip traces are evident in nearly every grain in the SEM micrographs presented in Figs. 3a and 4a, which cover most of the deformed gage length of the sample. From EBSD scans, the slip trace directions and Schmid factors (SF) were computed based upon the (convenient) assumption of uniaxial tension. The EBSD orientation maps are colored based upon the horizontal (**Y**) tensile axis inverse pole figure. Most of the grains are red-orange, indicating a near-cube orientation. Some of the other grains with other colors provide the possibility to study different kinds of grain boundaries. For example, the cyan grain serves to illustrate some important distinctions between slip in cube *vs.* non-cube orientations. To provide a general view about how the grains percolate through the thickness, the EBSD maps of the both front and back surfaces are shown for comparison in Figs. 3b and 3c, respectively. The map of the back surface (Fig. 3c) is presented as if the viewer was looking at the back surface from the inside out, which facilitates identification of grains that percolate through the entire thickness of the sample, indicating that about half of the grains on the front surface are also present on the back surface. For example, grain 17 is small on the front surface and much larger on the back surface. All of the EBSD data for the grain orientations and selected grain boundary misorientations are presented in the appendix in Tables A.1 and A.2. The overall appearance of the microstructure is similar for sample #4 and is not shown.

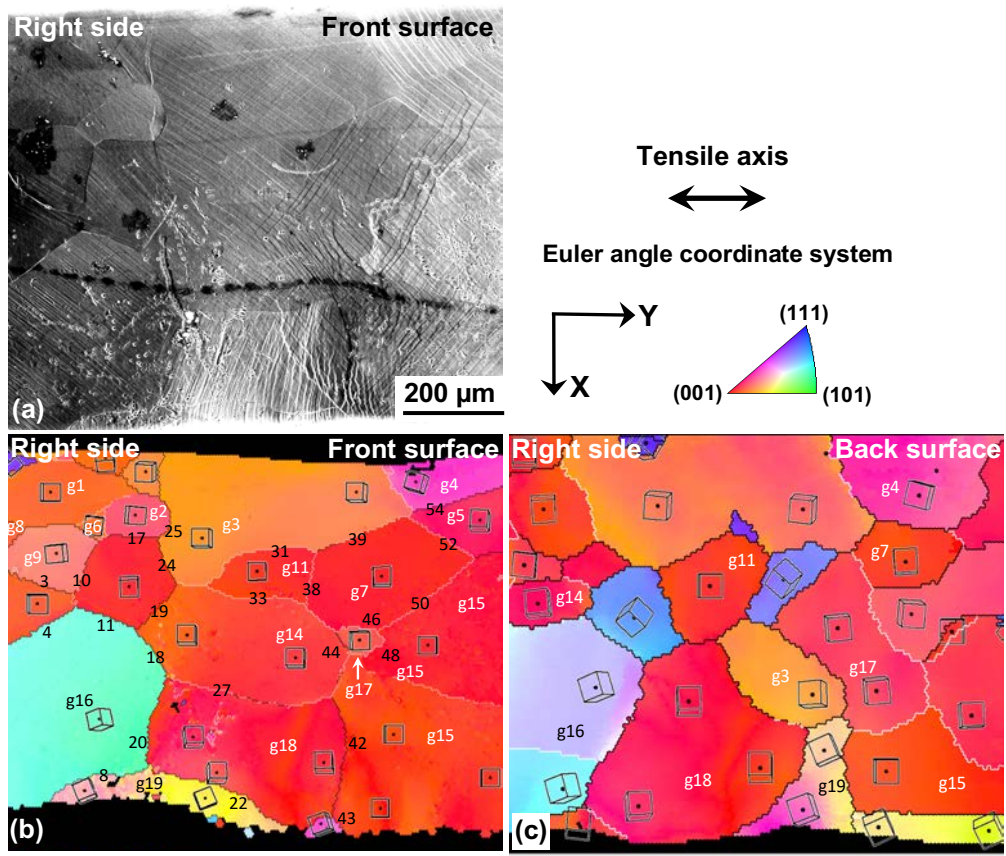

Fig. 3. (a) SEM micrograph of the right side of the deformed polycrystalline Al sample, showing slip traces in different grains. (b) Corresponding crystal orientation map (tensile axis inverse pole figure map) obtained by EBSD of the front surface of the sample. (c) Crystal orientations obtained by EBSD of the back surface of the right side of the sample (viewed in same direction as the front). Grains on the front surface and analyzed boundaries are labelled in white (mostly) and black fonts, respectively. High angle boundaries (>15°) are marked with black lines, and low angle boundaries between 6 and 15° are marked with white lines. In (a), irregular dark patches and the row of dark spots on the SEM image are surface contamination artifacts, and the EBSD maps are somewhat distorted due to heterogeneous surface topography, (but the orientation measurements are not affected).

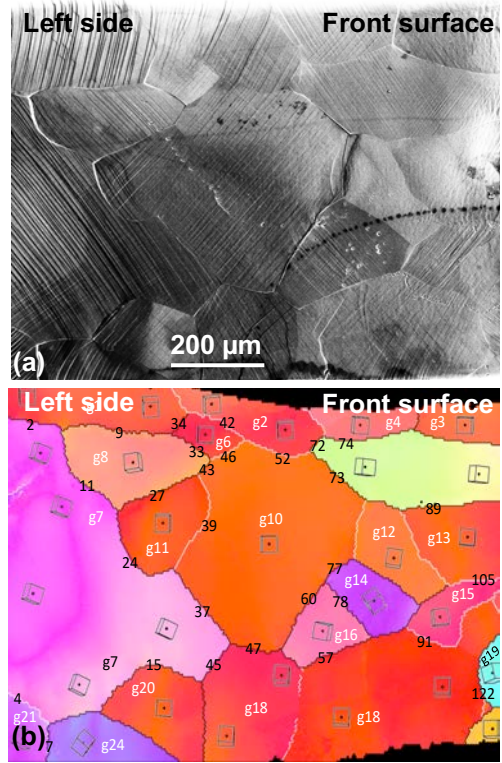

Fig. 4. (a) SEM micrograph of the left side of the deformed polycrystalline Al sample, showing slip traces in different grains. (b) Corresponding EBSD crystal orientation map, as in Fig. 3.

*3.1 Analysis considerations:*

In this paper, four colors are used to represent the four different FCC slip planes (each having three slip directions). The slip systems are identified in Table 1, where they are ordered based upon which quadrant the slip plane normal points in the crystal coordinate system. These slip systems are colored red, green, blue, and gold consistently throughout the paper. In the detailed images of slip traces that follow, starting with Fig. 5, the computed slip traces of all four slip planes are colored according to their {111} plane color. The crystal coordinate **x**, **y**, and **z** directions are indicated on unit cell prisms shown below the micrograph with red, green, and blue dotted lines along cube edges, respectively. For the cube orientations, the ideal cube-oriented grain deformed along a <100> direction in uniaxial tension has eight slip systems with a Schmid factor of 0.41 and four with a Schmid factor of 0. Among the three slip directions on each plane, the first one in each group (1, 4, 7 and 10, underlined in Table 1) has a Burgers vector nominally parallel to the surface with no **Z** component. These systems are likely to be active but they will not contribute much surface displacement to a slip trace due a small out-of-plane component of the Burgers vector (if the orientation deviates from the perfect cube orientation, they will contribute a very small surface displacement). Accordingly, they are referred to as 'invisible' slip systems. The second slip direction for each plane has a 0 **y**-component (in the crystal system). Because all of the near-cube grains have the crystal **x**-axis close to the vertical **X** direction, the second slip vector of each set is nearly perpendicular to the stress direction (**Y**), and consequently has a very low Schmid factor (slip systems 2, 5, 8, and 11). The third member of each set (3, 6, 9, 12) has a Burgers vector with a large out-of-plane component as well as a large component along the tensile **Y** axis, so these slip systems have the most visible slip traces. These trends are not relevant for the large cyan grain or other orientations that are not close to the cube orientation.

Table 1. Slip system definition; the same coloring method is used throughout the paper. The underlined systems (1,4,7 and 10) are those that could be invisible because the Burgers vectors of these systems are nearly parallel to the specimen surface

| Slip system | Plane | Direction |
|---|---|---|
| <u>1</u> | (111) | [$\bar{1}$10] |
| 2 | (111) | [10$\bar{1}$] |
| 3 | (111) | [0$\bar{1}$1] |
| <u>4</u> | ($\bar{1}$11) | [$\bar{1}\bar{1}$0] |
| 5 | ($\bar{1}$11) | [101] |
| 6 | ($\bar{1}$11) | [01$\bar{1}$] |
| <u>7</u> | ($\bar{1}\bar{1}$1) | [1$\bar{1}$0] |
| 8 | ($\bar{1}\bar{1}$1) | [$\bar{1}$0$\bar{1}$] |
| 9 | ($\bar{1}\bar{1}$1) | [011] |
| <u>10</u> | (1$\bar{1}$1) | [110] |
| 11 | (1$\bar{1}$1) | [$\bar{1}$01] |
| 12 | (1$\bar{1}$1) | [0$\bar{1}\bar{1}$] |

Fig. 3 clearly shows the heterogeneous deformation in the sample; a neck began to develop along the lower edge of the sample, near grains 18 and 19 on the right side, and these grains show more intensive slip traces than many other grains. Heterogeneous deformation leads to differences in the elevation of the initially flat surface, which affects the interpretation of slip traces. This surface elevation effect is illustrated in a more informative manner using several grains in the upper left side of the sample shown in Fig. 5. Slip systems associated with the dominant slip traces are illustrated using unit cells and slip plane triangles within them (representing a particular slip system) for several slip traces. Focusing on grain 1, there are two sets of slip traces near the boundary marked 2, some that are fairly closely aligned with the computed dashed blue plane trace. There are also very regularly spaced traces that curve to the left near the upper edge of the specimen, which are not very close to the computed gold plane slip trace. The gold and green slip planes for grain 1 are illustrated on a Thompson tetrahedron on the left side of grain 1, where the left and right sides (two slip planes) represent material that is under the surface, and the common edge is above the surface. On each face, the lines marked with a 'f' represent the ideal slip trace on a *flat* surface (the slip traces start at the lower left corner, and they diverge, indicating that the upper right edge of the tetrahedron protrudes above the sample surface). If the surface is *sloped*, such that the sample is thinner toward the top edge, then the slip traces would follow a path 's' that is to the left of the ideal trace for the gold plane, and to the right for the green plane. Because the green plane is more steeply sloped from the surface, the deviation from the ideal line is smaller. If the surface is *curved*, then the trace of the slip plane would follow a curved trajectory 'c' that increases in curvature as one approaches the edge of the sample. From this 3-dimensional virtual removal of material from the Thompson tetrahedron to reach the actual surface, it is apparent that the lack of agreement between the ideal gold trace and the observed trace is consistent with the curved surface. As there are no slip traces in grain 1 that are even close to the ideal green slip trace, there is no evidence for slip on the green plane.

A similar argument can be made regarding deviations from linearity of the slip traces for the blue plane in grain 7 to detect local surface topography. Just above the label for grain 7 (g7), there are deviations in the plane trace toward the left where there is an etched valley that is nominally perpendicular to the slip traces, which correlates with a low angle boundary evident in the upper left part of Fig. 3b.

In addition to being able to interpret deviations from agreement with an ideal slip trace, it is also possible to infer the sense of shear on a slip plane from the geometry of the step. The secondary electron detector is located above and to the 'northwest' of the sample, such that surface topography can be interpreted as if the detector was a light source. The grey slip plane triangles within the unit cells below this image defined by the slip systems in Table 1 have a plane normal with a -**Z** component (in other figures, the slip plane is a lighter tan color, indicating that the plane normal has a +**Z** component). The cyan vectors on one edge of the slip plane identify the Burgers vector direction starting from the dot end, on the positive face of the slip plane. In the two Thompson tetrahedrons, this sense of shear is illustrated with two vectors, one above and one below the slip plane for the gold and blue planes. This sense of shear is consistent with deformation in tension in the horizontal direction. A consequence of this direction of shear is that the material to the left of the trace is lower than the material to the right of the trace, resulting in a step that is in shadow with respect to the detector. Thus, the sense of shadow (or brightness) of a slip trace can also be used as a consistency check when slip traces may not be closely aligned with the ideal plane trace direction.

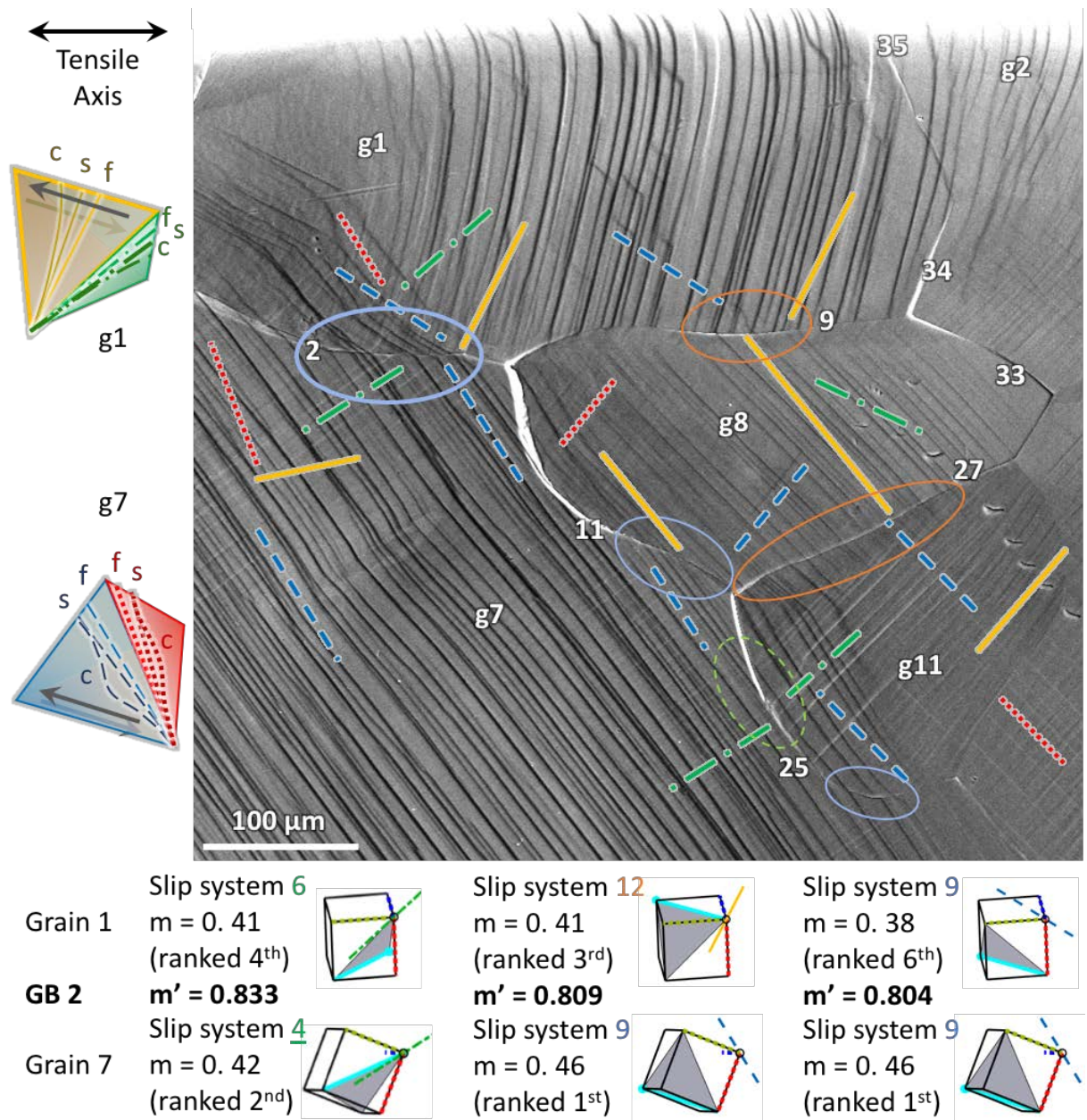

Fig. 5. (a) Several grains in the upper left side showing slip traces with curvature, a consequence of surface topography, which is discussed in the text using the Thompson tetrahedron for grains 1 and 7, where *c*, *s*, and *f* refer to surfaces that are curved, sloped, or flat, respectively. The ellipses identify instances of slip transfer, which are discussed and compared with other examples in the later part of the text. The prisms below illustrate slip system pairs with $m'$ values associated with observed slip transfer in grain boundary 2; their slip system plane trace colors (lines passing through the origin) are color coordinated.

The computed Schmid factors identify the likelihood that slip systems will be active, and they are ranked in tables of $m'$ values in order of decreasing Schmid factor for each grain sharing a grain boundary. In the analysis for grain boundary 4 (gb4) in Table 2, the left most column lists the slip system numbers for grain 12 (g12ss), and the next column to the right provides the corresponding Schmid factor for that slip system, and they are colored according to the slip plane. Similarly, neighboring grain 16 has slip system numbers across the top (g16ss), and Schmid factors below them in the next row. Because the *local* stress tensors are not known as a function of position anywhere in the sample, the ranking is *approximate*; when Schmid factors are close to each other, they have a similar likelihood to be active. The specific $m'$ value for each slip system pair is within the body of the table. This organization of $m'$ values based upon Schmid factors results in a 12 × 12 array with 144 $m'$ values, but the ranking by Schmid factor places the $m'$ values of greatest interest in the upper left corner of the table. For gb4 in Table 2, grain 16 has only four slip systems (ranked across the top) with Schmid factors greater than 0.25, but grain 12 has eight slip systems listed down the left side of the table with Schmid factors between 0.35 and 0.45. Therefore, in this boundary, there are 4 × 8 = 32 $m'$ values for slip systems that are likely to be active in this pair of grains. In the other tables presented below that focus on near-cube grain orientations as neighbors, the upper left part of an $m'$ table with high Schmid factors has 8 × 8 = 64 $m'$ values.

Table 2. Schmid factors, $m_i$, where $i$ denotes the rank in the descending list, for grains 12 and 16, and $m'$ values for grain boundary 4 on the right side. The slip traces of underlined systems are expected to be nearly invisible on the surface in near-cube oriented grains. The $m'$ values > 0.75 are highlighted in bold

| gb4 g12ss | g16ss $m_i$ | <u>4</u><br><u>0.46</u> | 5<br>0.40 | <u>1</u><br><u>0.35</u> | 2<br>**0.27** |
|---|---|---|---|---|---|
| <u>10</u> | <u>0.45</u> | 0.503 | 0.663 | -0.133 | -0.049 |
| 7 | 0.45 | -0.008 | 0.007 | **0.760** | 0.597 |
| 3 | 0.41 | 0.125 | 0.033 | 0.298 | -0.140 |
| 9 | 0.40 | 0.005 | -0.049 | 0.071 | 0.661 |
| 12 | 0.39 | -0.045 | 0.435 | 0.031 | 0.283 |
| 6 | 0.38 | **0.814** | 0.218 | 0.081 | -0.038 |
| <u>1</u> | <u>0.35</u> | 0.014 | -0.012 | 0.598 | 0.470 |
| <u>4</u> | <u>0.35</u> | 0.662 | **0.871** | -0.067 | -0.025 |

Although five slip systems are required for accomplishing an arbitrary shape change, fewer slip systems can also enable arbitrary shape changes in a grain neighborhood if a neighboring grain deforms in a compatible way. On the other hand, near-cube orientations have as many as 8 slip systems that could be activated, such that cube oriented grains can easily accommodate a neighboring grain that has far fewer slip systems that can be activated, as demonstrated in Table 2. One way of identifying a meaningful $m'$ value for a given boundary is to consider the upper left portion of an $m'$ table where slip systems have Schmid factors greater than a threshold value. In Table 2, $m'$ values that exceed a (chosen) threshold of 0.75 are in bold font, so that the value of 0.871 may be the most meaningful value of $m'$ among slip systems that have Schmid factors larger than 0.25. This approach provides an that enables comparison of the slip transfer potential among many boundaries. This maximum $m'$ value is represented with colored line segments in the grain boundary map in Fig. 6 for the front side of the sample, in which it is evident that there are many boundaries with high $m'$ values among grains with near-cube orientations. The $m'$ values of multiple segments of the same boundary have the same $m'$ value because the average grain orientation was used.

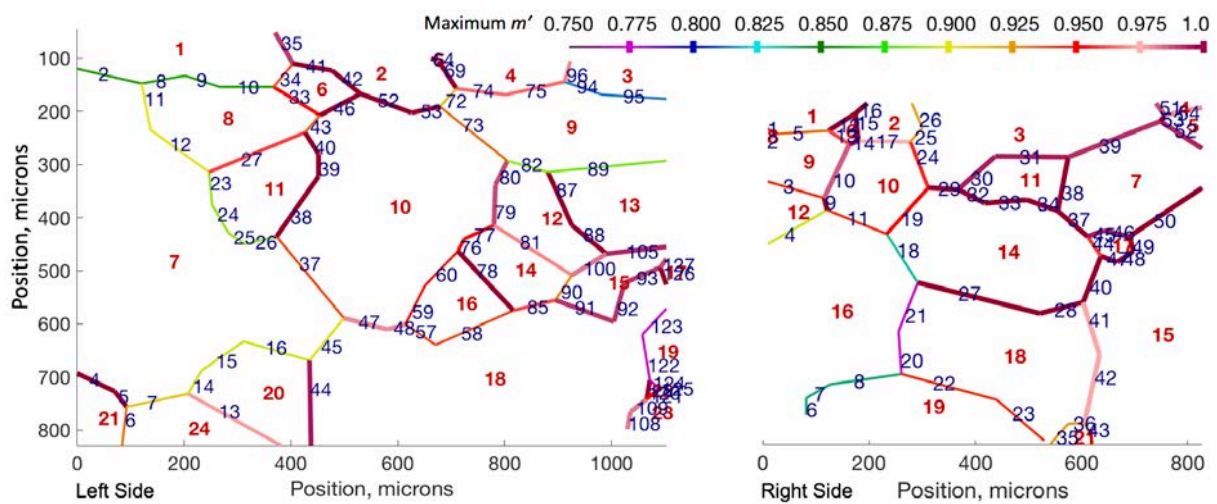

Fig. 6. A plot of maximum $m'$ values among slip systems with Schmid factors greater than 0.25 on grain boundaries on the front side of the left and right sides of the sample, according to the color scale at the top (the highest $m'$ values have thicker lines).

To analyze the possible activation of different slip systems in each grain, post-deformation SEM images of the sample are examined in detail to check for agreement between the observed slip traces in each grain and the computed slip traces of slip systems. In the sections that follow, grain boundaries that illustrate conditions where slip transfer was not observed, possibly observed (e.g. when a ledge is present at the grain boundary, indicating heterogeneous deformation on both sides, such that slip transfer may not be significant even if slip traces suggest that slip transfer occurred) and convincingly observed (when imposing and receiving slip traces in grains are clearly identified, and there is little topography along the grain boundary) are illustrated and analyzed. Most of the boundaries on the front of the sample (and on the front of sample #4) have been analysed, and a summary outcome of this analysis will conclude the paper.

*3.2 Slip transfer was not observed*

As indicated above, Table 2 provides $m'$ values for grain 12 (a red near-cube orientation) and grain 16 (the cyan orientation), which share grain boundary 4. In the near-cube oriented grain 12, one pair of slip systems have Schmid factors $m_i$ higher than the ideal 0.41 value (slip systems 10 and 7 have $m_1 = 0.454$ and $m_2 = 0.451$, where the subscript *i* indicates the rank in the Schmid factor list). The next pair has Schmid factors near the ideal cube value (systems 3 and 9), and two other pairs have values lower than the ideal cube orientation (systems 12, 6, 1 and 4). The four omitted slip systems have Schmid factor values below 0.07. In contrast, only four slip systems (systems 4, 5, 1 and 2) in grain 16 have Schmid factors above 0.25. Each element of the array has the $m'$ value for the slip systems in the corresponding column and row. There are only three $m'$ values above 0.75 (bold in Table 2), which are associated with the three most highly stressed slip systems in the cyan grain 16 (top row), but the two above 0.80 correspond with lower Schmid factor slip systems in the red grain 12 (left column), and one of them has the highest $m'$ value of 0.871. Accordingly, boundary 4 on the left edge of the right side of the specimen has a light green color based upon the color scale (Fig. 6b).

The secondary electron micrograph in Fig. 7 shows slip traces in grains 12 and 16 near grain boundary 4. The potential slip trace directions corresponding to all four slip planes are shown in both grains, in four different colors according to the colors presented in Table 1. Considering the Schmid factors in each grain, and checking the correspondence of the slip traces in each grain with the computed slip trace lines, it is possible to identify the dominant slip systems in each grain. For example, there are slip traces that are nearly aligned with the green slip trace, and faint evidence for activation of slip systems on the red or blue planes on the left edge of red grain 12. The two slip systems with the highest Schmid factors are for 'invisible' slip systems on the gold and blue planes (Table 2), so the faint evidence for slip on the blue plane is consistent with the calculation. The slip traces close to the green plane trace are more evident, and the mostly highly stressed system on a green plane is for a visible system, but it has the 6th highest Schmid factor. Below the micrograph in Fig. 7, the slip systems corresponding to the two highest $m'$ values are illustrated, which involve two different slip directions on the same (green) planes in both grains. Slip systems 4 and 5 in cyan grain 16, which have the highest Schmid factors, correspond with slip systems 6 and 4 in the near-cube grain 12, which have respectively the 6th and 8th highest values of the Schmid factor. (The slip direction in system 4 is 'invisible', so the observed slip traces in grain 12 are probably due to the other slip direction for system 6, which has a slightly higher Schmid factor). By tracking the green slip traces in both grains toward the grain boundary in Fig. 7(a), it is apparent that even though the slip traces are fairly closely aligned with each other, the traces near the boundary almost disappear, suggesting that dislocations were repelled by the boundary and that slip transfer did not happen in this case. Also, a ledge developed at the boundary, which further indicates that the two grains deformed heterogeneously.

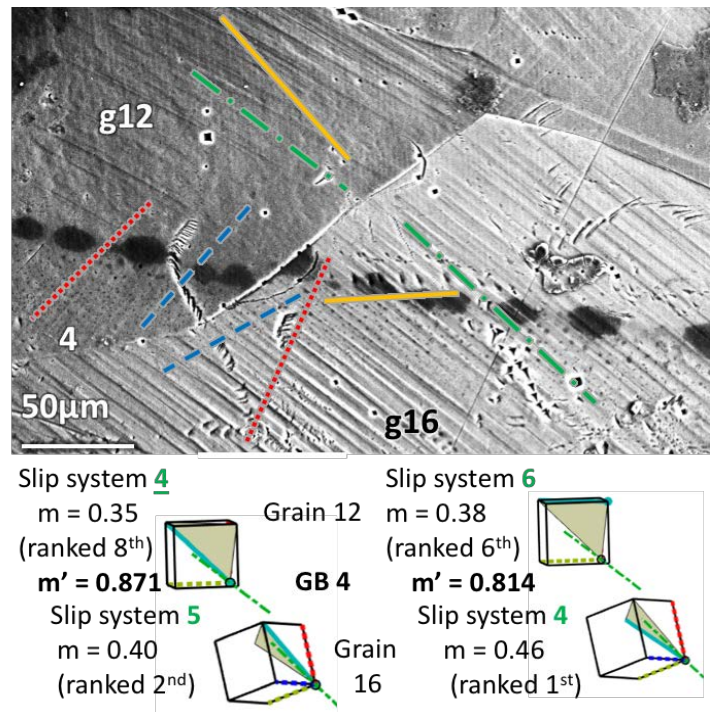

Fig. 7. (a) Secondary electron image showing the slip traces at grain boundary 4 on the right side, with a misorientation of 48.7° between grains 12 and 16. (b) The Schmid factors in grain 12 that have high $m'$ values with grain 16 have low Schmid factors, and because the slip traces are not present at the boundary and there is a ledge, slip transfer was unlikely.

Grain boundary 9 on the left side of the sample illustrated in Fig. 5 also shows resistance to slip transfer. In this case, slip traces in grain 1 stop abruptly at the grain boundary. The surface of grain 8 is nearly smooth adjacent to the boundary 9 excepting for a couple traces. Further below the boundary, there are many slip traces on the gold slip plane, suggesting that the boundary resisted dislocation approach. There are two possible instances of slip transfer to the blue slip system in grain 1, between the number '9' and two of the largest ledges on the gold plane in grain 8, and aligned with intermittent instances of cross slip from the gold plane to the blue plane in grain 1. However, this is exceptional, and overall the boundary resisted slip transfer. The boundary is also twisted; the elevation of grain 8 is below grain 1 on the left side, and above grain 1 in the middle, indicating that strains in the two grains differed spatially along the boundary. Accordingly, taking into account these two observations from grain boundaries 4 and 9, this indicates that (consistent) slip transfer requires a $m'$ value greater than 0.87.

*3.3 Slip transfer appears to have happened, but did not or may be uncertain*
Aligned slip traces were often found across the boundary between two grains with near-cube orientation, such as those in Figs. 8 and 9. In boundary 25 between grains 2 and 3 in Fig. 8, the apparent alignment is between a green trace and a gold trace, but inspection of the aligned slip systems shows that the slip directions and the plane normal directions are highly misaligned. There are high values of $m'$ between both of the green and gold slip trace pairs, but the *observed* slip traces are not for the same kind (color) of compatible slip system. Also, in the case of grain boundary 10 between grains 10 and 9 (Fig. 9), the slip traces in grain 10 could be from either the green or gold slip traces, but because the slip step is bright (inclined toward the detector), it is more consistent with the compatible slip system both gold planes. Nevertheless, a ledge developed at both boundaries, indicating heterogeneous strain in the two grains, such that slip transfer was not a dominant effect.

The corresponding table of $m'$ values for the slip systems with the highest Schmid factor for grain boundaries 25 and 10 are presented in Tables 3 and 4, respectively. The $m'$ in Tables 3 and 4 show that the high $m'$ values in near-cube oriented grains are between the same slip system in the neighbor grain for all eight slip systems and, thus, slip transfer is expected between same-color slip systems in both grains. However, the $m'$ value of 0.071 circled in Table 3 for the apparently aligned slip traces in boundary 25 (the aligned green and gold plane traces) is very small.

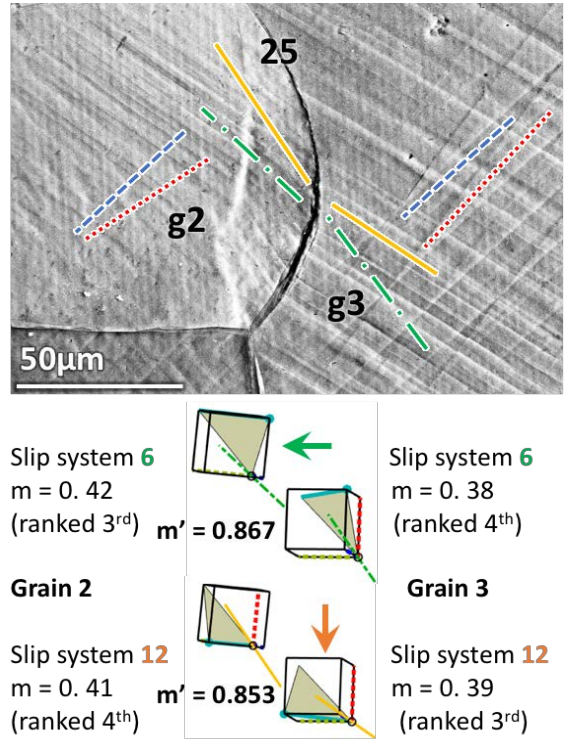

Fig. 8. (a) Secondary electron image showing the slip traces at grain boundary 25 on the right side, with a misorientation of 26.2° between grains 2 and 3. The apparent common traces are on green planes in grain 2 and gold planes in grain 3 (arrows), which are not compatible for slip transfer.

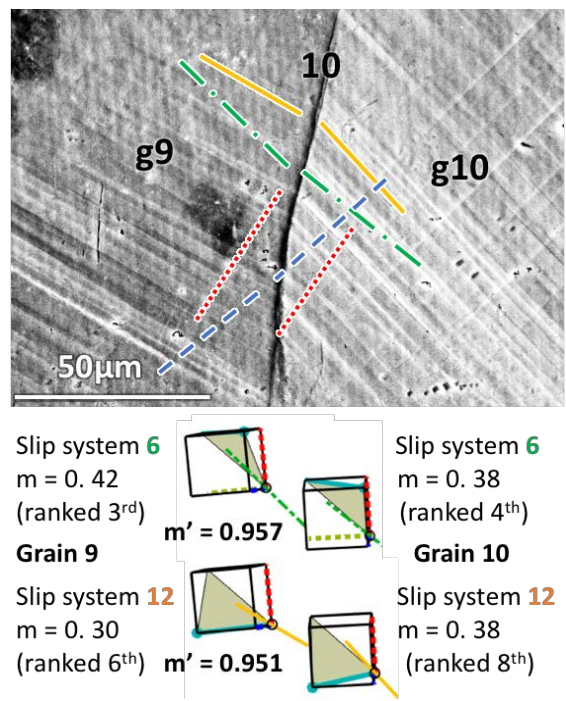

Fig. 9. (a) Secondary electron image showing the slip traces at grain boundary 10 on the right side, with a misorientation of 16.0° between grains 9 and 10. Though gold traces appear to be correlated at the boundary, they are less favored, and there is a large ledge, indicating that the two grain deformed heterogeneously.

Table 3. Schmid factors *m* for grains 2 and 3 and $m'$ values for grain boundary 25. The slip traces of underlined systems are nearly invisible on the surface. The apparent $m'$ value for aligned slip systems in Fig. 8 is very low (bold light green value).

| gb25 g2ss | g3ss $m_i$ | 1 0.48 | 4 0.47 | 12 0.39 | 6 0.38 | 3 0.36 | 9 0.35 | 10 0.30 | 7 0.30 |
|---|---|---|---|---|---|---|---|---|---|
| 7 | 0.47 | 0.665 | -0.003 | 0.087 | -0.037 | 0.242 | 0.595 | 0.003 | 0.898 |
| 10 | 0.45 | 0.000 | 0.490 | 0.636 | 0.072 | 0.000 | 0.276 | 0.862 | -0.043 |
| 6 | 0.42 | 0.128 | 0.726 | 0.071 | 0.867 | 0.204 | -0.237 | 0.128 | -0.320 |
| 12 | 0.41 | 0.001 | 0.102 | 0.853 | -0.219 | -0.001 | 0.370 | 0.179 | 0.126 |
| 9 | 0.34 | 0.210 | -0.020 | 0.122 | 0.043 | -0.278 | 0.834 | 0.025 | 0.284 |
| 3 | 0.32 | 0.510 | 0.245 | -0.265 | 0.293 | 0.814 | -0.024 | -0.478 | -0.032 |
| 4 | 0.32 | -0.024 | 0.885 | 0.115 | 0.130 | 0.031 | -0.385 | 0.156 | 0.060 |
| 1 | 0.30 | 0.885 | 0.008 | -0.412 | 0.116 | 0.322 | -0.037 | -0.016 | -0.055 |

Table 4. Schmid factors *m* for grains 9 and 10 and $m'$ values for grain boundary 10. The slip traces of underlined systems are invisible in the surface. The highest $m'$ values for observed slip systems have low Schmid factors (circled).

| gb10 g9ss | g10ss $m_i$ | 9 0.43 | 3 0.43 | 4 0.42 | 1 0.42 | 7 0.39 | 10 0.39 | 6 0.38 | 12 0.38 |
|---|---|---|---|---|---|---|---|---|---|
| 4 | 0.49 | -0.071 | 0.262 | 0.956 | 0.020 | -0.012 | 0.381 | 0.690 | 0.127 |
| 1 | 0.47 | 0.245 | 0.479 | -0.018 | 0.958 | 0.567 | 0.009 | 0.130 | -0.057 |
| 9 | 0.40 | 0.938 | 0.017 | -0.278 | 0.042 | 0.550 | 0.285 | -0.101 | 0.439 |
| 3 | 0.38 | -0.125 | 0.926 | 0.069 | 0.461 | 0.273 | -0.035 | 0.252 | 0.029 |
| 6 | 0.32 | 0.049 | 0.363 | 0.261 | 0.181 | -0.106 | 0.104 | 0.957 | -0.087 |
| 12 | 0.30 | 0.170 | -0.117 | 0.184 | -0.292 | 0.100 | 0.618 | 0.067 | 0.951 |
| 7 | 0.28 | 0.416 | 0.036 | 0.030 | 0.073 | 0.961 | -0.031 | -0.219 | 0.194 |
| 10 | 0.26 | 0.056 | -0.356 | 0.280 | -0.027 | 0.009 | 0.942 | 0.202 | 0.314 |

Additional insights can be obtained from both boundaries by considering visible and invisible slip systems in Table 3. The highest two Schmid factors in both grains 2 and 3 along grain boundary 25 are for invisible slip systems (7 and 10 in grain 2 and 1 and 4 in grain 3), but for each pair, the corresponding slip systems in the other grain have the lowest Schmid factors (Table 3). The next highest Schmid factor slip systems are visible systems 6 and 12 in grains 2 and 3, but only one of the two is evident in each grain, and for a pair that is incompatible. Given that no slip transfer between this pair was convincing (the significant ledge in the boundary indicates significant differences in how the two grains deformed), this also suggests that slip transfer is not favored with an $m'$ value below 0.9. Also, use of local orientations at the boundary (rather than average grain orientations), made no difference in the relative ranking of Schmid factors, and $m'$ values in the table only slightly different (not shown).

In contrast, the $m'$ values for boundary 10 (Table 4 and Fig. 9) are all well above 0.9. The two highest Schmid factors in grain 9 are for *invisible* slip systems (slip systems 4 and 1) with $m'$ > 0.95, coupled with the third and fourth highest Schmid factor slip systems in grain 10. These numbers suggest that although slip transfer with $m'$ > 0.95 may has occurred, it cannot be confirmed by slip traces because the slip systems are nearly invisible. The two highest Schmid factor slip systems in grain 10 are the visible red and blue systems (9 and 3), and while there is some evidence for the blue slip system some distance from the grain boundary in grain 10, no evidence of activation of this system can be found in grain 9. The observed green and gold slip traces are for slip systems 6 and 12 which have much lower Schmid factors, suggesting that the stress state in these grains was significantly different from the assumed uniaxial tension, which would make operation of the invisible systems 1 and 4 was less likely. These considerations along with the apparent ledge in the grain boundary indicate that heterogeneous slip took place in the two grains, indicating that even $m'$ values near 0.95 may not be sufficient to cause significant slip transfer.

*3.4 Slip transfer probably occurred*

Fig. 10 and Table 5 show an example of probable slip transfer at grain boundary 52 from grain 7 to grain 5. The slip traces aligned with the gold slip systems in both grains are bright, consistent with the slip plane being inclined toward the detector (note that the crystal **x** axes in Figs. 5 and 10 point in opposite directions; the crystal coordinate systems are rotated approximately 180° about the horizontal **Y** axis). In contrast, neither the red or blue slip traces with $m'$ values of 0.962 and 0.945 correspond closely with the observed shadowed slip traces. These slip traces are contiguous across the boundary (which is located under the '52'), but the traces are displaced to the left in grain 5, suggesting that the red slip traces occurred prior to subsequent deformation on the gold slip systems. These dark slip traces are long – they are evident in the upper half of the right side of Fig. 3a, and they are continuous across grains 7, 14, 15 and 5, but they are not straight, suggesting that they occurred early in the deformation process and became distorted by later slip on other systems. The dark shadowed traces imply that the traces are more consistent with the red slip plane. Given that this boundary is near the top of the sample, where a surface slope is likely (similar to Fig. 5) the observed slip traces on the red plane deviate to the left, consistent with a lower elevation toward the top. The development of the sloped surface can be understood from the sense of shear taking place on the gold and red slip planes: the gold system will move material down and to the right, while the red system will move material down and to the left, leading to a significantly reduced elevation in the neighborhood. The red system has a higher Schmid factor in grain 7 (Table 5) while the gold system has a higher Schmid factor in grain 5, and the amount of slip displacement corresponds with this difference, as the red traces are arrested in grain 5 and the surface steps on gold planes are smaller in grain 7.

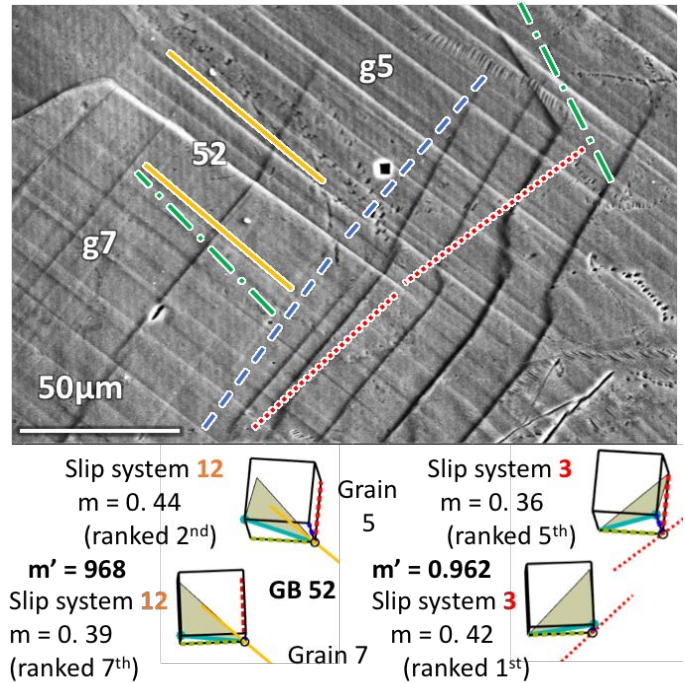

Fig. 10. (a) Secondary electron image showing slip traces at grain boundary 52 on the right side, with a misorienation of 14.4° between grains 5 and 7 The illuminated slip traces parallel to the boundary are consistent with the gold slip plane, and the dark shadowed slip traces crossing grain boundary 52 are consistent with the red slip plane (rarer illuminated traces are consistent with the blue plane). The disagreement with the red slip traces are consistent with thinning of the sample toward the top edge.

Table 5. Schmid factors $m$ for grains 7 and 5 and $m'$ values for grain boundary 52 with a 14.4° misorientation. The most highly stressed four slip systems have low $m'$ values, but the red 3 slip systems exhibited slip transfer with a high $m'$ value but a lower Schmid factor in the receiving grain 5 where slip was arrested in Fig. 10.

| gb52 g7ss | g5ss $m_i$ | 1 0.44 | 12 0.44 | 4 0.43 | 6 0.43 | 3 0.36 | 10 0.36 | 9 0.36 | 7 0.35 |
|---|---|---|---|---|---|---|---|---|---|
| 3 | 0.42 | 0.526 | -0.040 | 0.300 | 0.519 | **0.962** | -0.185 | 0.015 | 0.067 |
| 9 | 0.42 | 0.300 | 0.294 | -0.042 | 0.018 | -0.060 | 0.083 | **0.945** | 0.565 |
| 10 | 0.42 | -0.044 | 0.696 | 0.289 | 0.120 | -0.134 | **0.969** | 0.257 | 0.048 |
| 7 | 0.41 | 0.513 | 0.120 | 0.019 | -0.072 | 0.244 | -0.037 | 0.387 | **0.966** |
| 1 | 0.40 | **0.961** | -0.128 | -0.064 | 0.247 | 0.457 | 0.040 | 0.049 | 0.122 |
| 4 | 0.40 | 0.016 | 0.259 | **0.944** | 0.393 | 0.048 | 0.361 | -0.336 | -0.063 |
| 12 | 0.39 | -0.193 | **0.968** | 0.081 | -0.035 | 0.039 | 0.273 | 0.358 | 0.214 |
| 6 | 0.39 | 0.065 | 0.046 | 0.558 | **0.966** | 0.118 | 0.213 | -0.060 | -0.262 |

Fig. 11 shows a grain on the left side, where there is a parent grain orientation (grain 16) with an annealing twin (grain 14), with possible slip transfer between grain 16 and neighboring grain 10. The operation of the highly favored slip system on the red twin plane is evident on both sides of the twin boundary (and possibly in the twin boundary), as the red traces have a light appearance consistent with the inclination of the twin plane. The $m'$ values for all three slip directions on the red plane are 0.999 in the twin boundary, but by definition, slip transfer does not occur on slip systems parallel to the boundary (instead this would be grain boundary sliding, and there is no evidence for preferential slip on the twin boundary). The lower half of the grain (parent orientation) also has slip traces parallel to the gold plane, which are not present in the upper half (twin), indicating that the two parts of the grain deformed differently (note the circled region at the triple point). Slip traces exist in neighbor grain 10 that are correlated with both the red and gold slip planes in g16, which have high $m'$ values and Schmid factors. However, the large ledge at the boundary makes the likelihood of significant slip transfer uncertain.

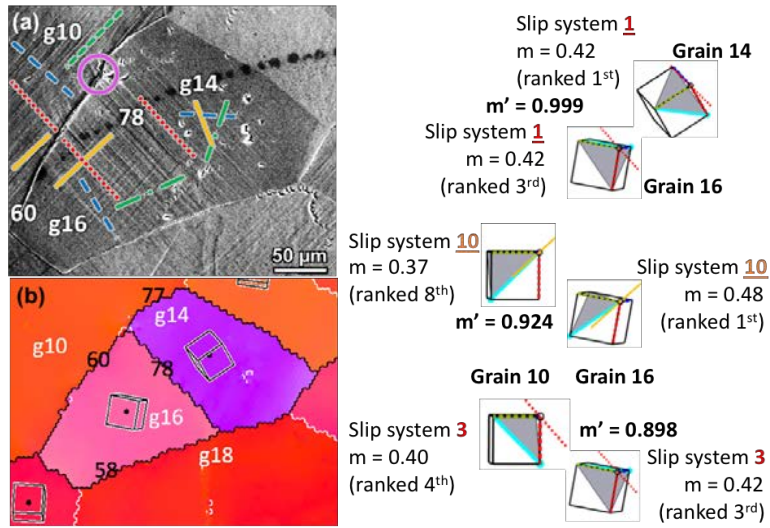

Fig. 11. (a) Secondary electron image showing the slip traces at grain boundaries 78 (between grains 14 and 16) and 60 (between grain 10 and 16) on the left side. (b) The purple ring and the EBSD inverse pole figure map show the location of grain boundary 78. Though $m'$ for boundary 78 is 0.999, the slip direction is parallel to the twin boundary, and does not represent slip transfer, which is more likely on grain boundary 60 on slip systems 3 and 10.

*3.5 Slip transfer occurred convincingly*

In Fig. 12, slip transfer is apparent across grain boundary 4 with the green slip systems in grains 7 and 21, which has an 8.6° misorientation. There are similarly high Schmid factors on the blue slip system in both grains. The $m'$ values for all of the slip systems are similarly high, ranging from 0.978 to .989, but the Schmid factors in both grains decrease almost as quickly as those for grain 16 in Table 2, so the third and fourth highest Schmid factors are not strongly favored. There is strong evidence for slip transfer through this boundary where only four slip systems have high Schmid factors and only three are evident (the one that is not evident, $m_2$ in grain 21, has the lowest Schmid factor). The slip trace topography on the green plane is greater in grain 21 than in grain 7, which has a lower Schmid factor along with the Burgers vector being closer to the surface in the receiving grain 7, which can account for the lesser topography of slip in grain 7. It is not clear why slip transfer did not occur in from grain 7 to grain 21, when slip traces were not repelled at the grain boundary.

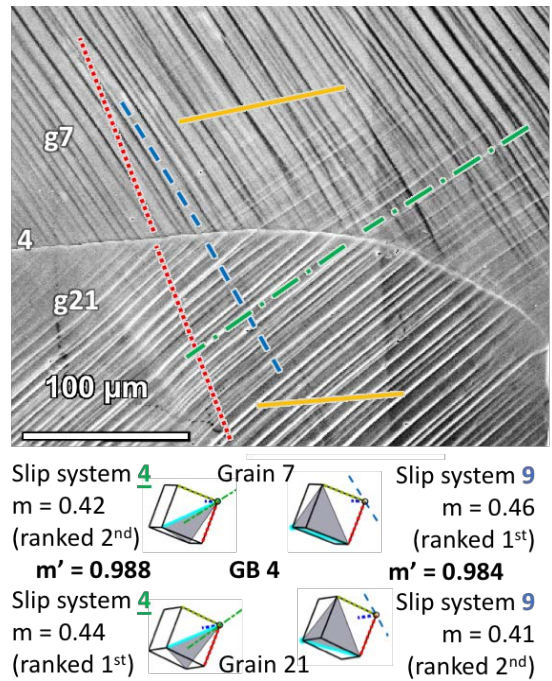

Fig. 12. (a) Secondary electron image showing the slip traces at grain boundary 4 on the left side, with a misorientation of 8.6° between grains 7 and 21. The high $m'$ values for the nearly invisible slip system 4 and a small ledge at the boundary indicate convincing slip transfer on the green slip system.

Fig. 13 shows a convincing example of slip transfer with a higher angle misorientation of 17.1° in comparison with grain boundary 4 indicated above (with a misorientation angle of 8.6°), where the gold visible slip system passed through a boundary with $m' = 0.954$, with no ledge topography along the grain boundary. The red invisible slip system has nearly the same $m'$ value, but there is no evidence for this slip system. The two slip directions on the green plane have lower $m'$ values, and lower Schmid factors, and no evidence of slip transfer, though there is some indication of activity of the visible green slip system in grain 5 below the grain boundary. Many of the slip traces appear to have originated in grain 4, as the steps are larger than their continuation in grain 5, and they spread out as they approach the boundary, with the exception of the trace just below the computed gold slip traces, where the trace has a larger step in grain 5.

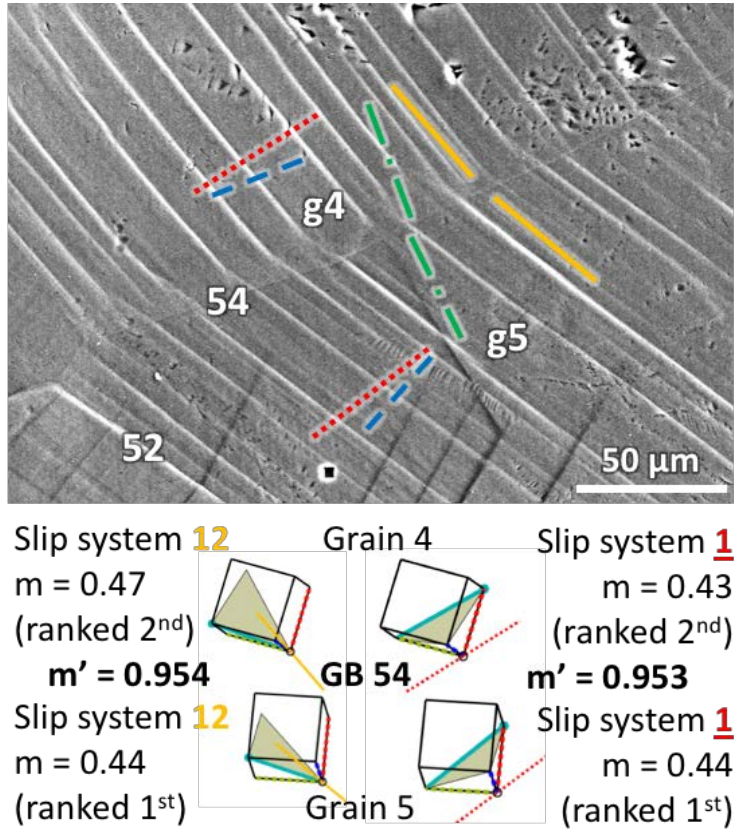

Fig. 13. (a) Secondary electron image showing the slip traces at grain boundary 54 on the left side, with a misorientation of 17.1° between grains 4 and 5. Though the red and gold systems have similar $m'$ values and high Schmid factors the gold system shows evident slip transfer (the red system is invisible, and the boundary has no ledge).

Grain boundary slip transfer is also evident in five boundaries identified by solid light blue and orange ellipses in Fig. 5 (grain boundaries 2, 9, 11, 25 and 27), which have misorientation angles between 21 and 42°, with small or no ledges in the boundary where the slip transfer across the grain boundaries took place. In this region, grains 1, 8 and 11 are adjacent to the very large grain 7, in which the visible blue slip system with the highest Schmid factor of 0.46 is dominant. This dominant blue slip system apparently initiated slip transfer across boundaries into compatible slip systems at grain boundaries 2, 11 and 25 in regions identified by light blue ellipses ($m'$ values of 0.804, 0.875, and 0.858, respectively; this directionality is inferred because the transmitted slip bands in grain 8 are less distinct). The apparent slip transfer between the blue slip systems at boundary 2 is lower than for two other slip system pairs as illustrated in the prisms below Fig. 5, so it is noteworthy that this transfer event happened. The next highest Schmid factor in grain 7 is 0.42 on the nearly invisible green slip system, which may be associated with slip transfer from grains 1 and 11 (with $m'$ values of 0.833 and 0.864, respectively; the $m'$ value from the green slip system in grain 7 to the compatible blue system in grain 11 is 0.762, and there is no evidence of slip transfer between these two systems). It is also possible that the green slip system in grain 7 may result from self-accommodation in response to heterogeneous strains in grains 1, 8, and 11. Because a ledge developed where slip transfer is apparent in the dashed green ellipse at grain boundary 25, self-accommodation by the green slip system in grain 7 is more likely than slip transfer. The orange ellipses identify instances of slip transfer of the gold slip system from grain 8 into grain 1 with an $m'$ of 0.835 (one or two instances). In contrast, there are many instances of slip transfer from the gold system into grain 11 with an $m'$ of 0.934, but on the left side of the boundary near grain 7, the direction of transfer appears to be reversed, because the less distinct traces are more likely in the receiving grain. Because grain 7 is large and not a cube orientation, and because only one slip system is highly favored, neighboring grains are constrained to accommodate a shape change imposed by this slip system, leading to conditions that appear to be more favorable for slip transfer than among the near-cube orientations examined elsewhere in the sample.

*3.5 Trends from over 128 observations of individual grain boundaries*

Fig. 14a shows how observations such as those reviewed above compare for 128 boundaries examined in samples 1 and 4. In this figure, the $m'$ value of the *observed* slip system pair ($m'_{Obs.}$) is plotted against misorientation angle of the boundary as solid red circles when the evidence is convincing for slip transfer, and as red circles with pink fill when the evidence is possible but not convincing (e.g. a ledge is present at the grain boundary). In each case, the corresponding maximum $m'$ ($m'_{Max.}$) is also plotted in blue squares in a similar manner. In cases where there was no evidence of slip transfer, the maximum $m'$ ($m'_{Max.}$) values are plotted as gray × symbols (specifically, the maximum $m'$ value for slip systems with Schmid factors greater than 0.25). It is notable that the $m'$ values follow a very well-defined trend of decreasing $m'$ value with increasing misorientation angle up to about 40°. In some cases, the observed slip transfer occurred with an m' that was slightly below the maximum $m'$ trend. There was more often uncertainty about significant slip transfer with misorientations above 14°, yet there were also some instances of convincing slip transfer up to about 36° (grain boundary 2). However, these observations at higher misorientations are all associated with grain 7 in Fig. 5, where the dominant grain was not cube-oriented. There are many cases where no slip transfer was observed along this trend line (gray × symbols); one such 11.5° misoriented boundary has $m' = 0.98$, and there are six instances for $m' > 0.95$. Slip transfer for near-cube orientated grains is likely when $m' > 0.97$, which implies that slip transfer is consistently favored only for low angle grain boundaries. This threshold value was defined as the minimum value of $m'$ in all studied cases which showed clear slip transfer in near cube oriented grains.

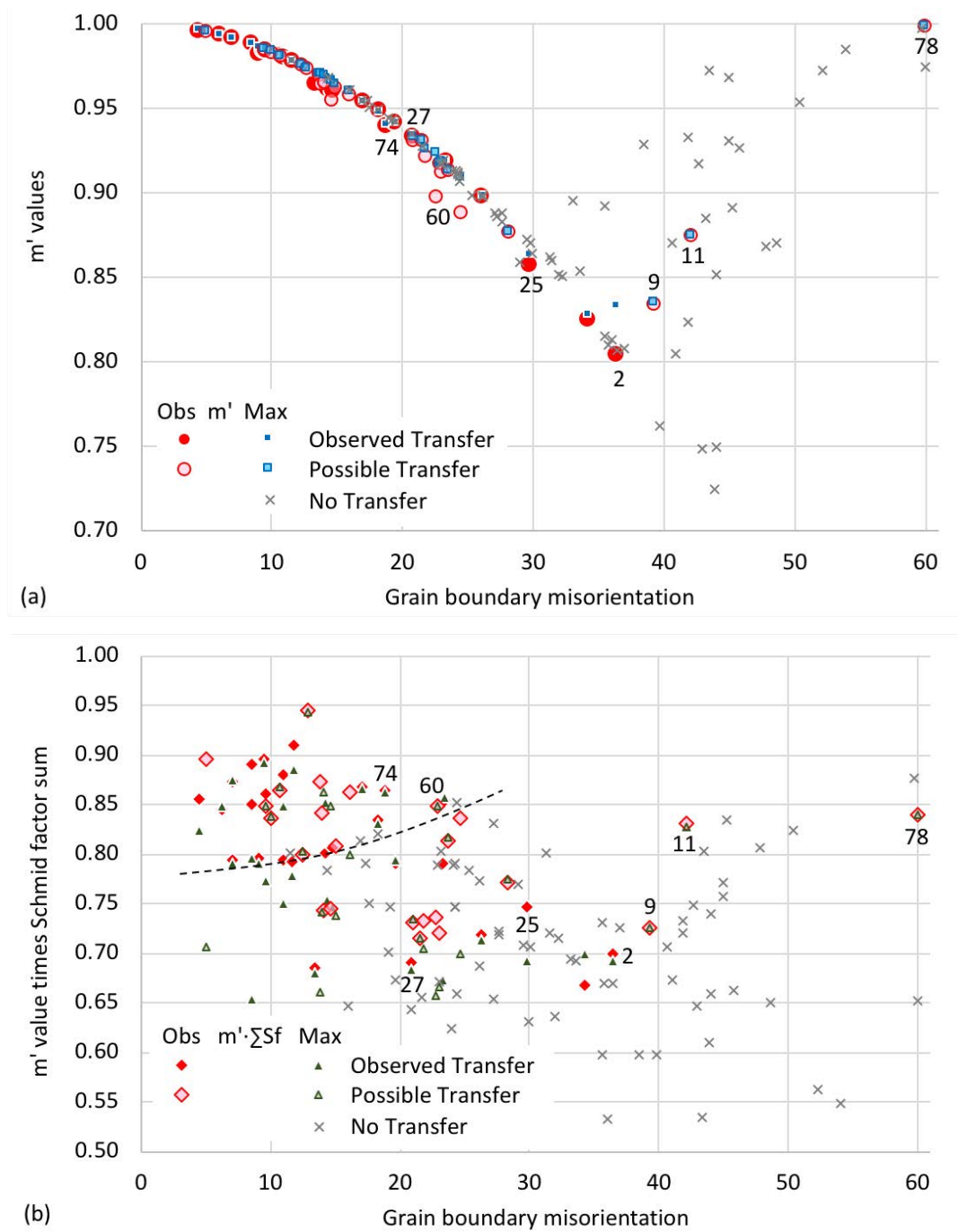

Fig. 14. (a) The relationship between $m'$ values and misorientation angle indicate that low angle boundaries always exhibit slip transfer, but with higher misorientation angles, slip transfer was only occasionally observed. (b) The product of $m'$ times the Schmid factor sum suggests that a threshold for likely slip transmission may exist for near-cube orientated grains as indicated by the dashed line. Numbered symbols are from boundaries associated with the large grain 7 non-cube oriented grain (Fig. 5), or the twinned grain in Fig. 11.

These results also indicate that the Schmid factor should be high in addition to the $m'$ value for slip transfer to be likely between two slip systems in neighboring grains. The effect of the Schmid factor is considered in Fig. 14b, where the product of $m'$ times the sum of the Schmid factors of the two slip systems (ΣSf) is plotted against misorientation angle in a similar manner as in Fig. 14a (using diamonds and triangles to distinguish ΣSf from $m'$). This figure shows a dashed line threshold below which no slip transfer was consistently observed (gray × symbols), but instances of slip transfer or possible slip transfer are occasionally observed. However, this trend is not followed by the numbered observations related to the large non-cube oriented grain 7 on the left side, suggesting that the criterion for higher angle boundary slip transfer for non-cube orientations is much more specific (the three unlabeled instances of convincing slip transfer from specimen 4 are from either large grains or larger misorientations).

4.  **Discussion**

Slip transfer in near-cube oriented grains appears to be less likely than in hexagonal materials, as slip transfer occurred consistently only for low angle boundaries. Above about 14° ($m' < 0.97$), instances of no slip transfer were increasingly observed. In boundaries with $m'$ values near 0.95, there was sometimes convincing evidence for slip transfer (Fig. 13 gold-gold), and sometimes not (a ledge in Fig. 9, boundary 10). It is also possible that the appearance of slip transfer is due to mutual and typically unequal generation of dislocations on incompatible slip systems that have similar slip traces (Fig. 8 in boundary 25). As these boundaries often had a distinct ledge, this implies that different amounts of slip occurred on either side of the boundary, so significant slip transfer through a transparent boundary that enables a more uniform strain did not happen. The boundaries that were least visible due to a lack of a ledge were those in which the most convincing slip transfer was observed, e.g. the gold traces in Fig. 13 and the portions of boundaries with ellipses in Fig. 5. The presence of grain boundary ledges was also correlated with the lack of slip transfer in pure tantalum (Bieler et al., 2014).

One possible reason for the lack of slip transfer is that grain boundaries could be made up of multiple kinds of dislocations that provide barriers to dislocations to pass through a grain boundary. This is a likely scenario because few of the boundaries investigated is a special boundary. The dislocations that make up non-special boundaries are probably interlocked through junctions such that they are unable to break out even when dislocations are piled up against the boundary. This is scenario was examined by means of dislocation dynamics simulations (Liu et al., 2012) but the analysis was done for a very low angle boundary (technically no misorientation, but only a network of dislocations was present).

Another potential reason why slip transfer was not commonly observed is that near-cube grains have eight slip systems with high Schmid factors and hence, high resolved shear stress. Hence, it may be easier to activate various intragranular slip systems (to accommodate the grain shape to the evolving boundary conditions imposed by neighbor grains) than to promote slip transfer across the boundary. This possibility is not generally available for hexagonal materials and, as a result, slip transfer is observed more often in higher misoriented boundaries (Bieler et al., 2014; Hémery et al., 2018). There is also evidence of sequential operation of slip systems in several places, such as the slip traces that are displaced within grain 5 in Fig. 10 (the displacements are not at the grain boundary, where the number '52' is located). Consideration of the effects of a sloped surface and 3-dimensional visualization of intermittent sequential operation of two slip systems that cross each other is necessary to account for features that do not agree with simple assumptions.

The fact that half of the slip systems expected to be active are 'invisible', i.e. the Burgers vector is nearly parallel to the surface, is a complicating factor for interpretation. Operation of only these slip systems would not generate significant surface topography, but they could cause a shear of slip traces from other systems that could confuse interpretation (in addition to effects of surface elevation), though there is no clear example of this possibility. Nevertheless, if slip in two slip directions on the same plane have high Schmid factors, even a little bit of slip on the 'visible' out-of-surface slip direction on the same plane would contribute to a visible slip trace. Also, activity of an invisible slip system in a grain that is tilted such that there is some out-of-plane component of shear displacement for an invisible slip system would make it faintly visible; examples of this are the green slip system in grain 7 in Figs. 5 and 12. Given that slip in FCC metals can be described as 'card glide', dislocations can easily move in two directions on the same plane, such that slip traces provide evidence of both slip systems. Consequently, slip system activation of 'invisible' slip systems is likely to be greater than the magnitude of the often faintly observed slip steps, such that faint slip traces may represent significant activity. High resolution digital image correlation measurements could confirm this hypothesis using an analysis strategy proposed by Chen and Daly (2017).

Given the unknown local stress tensor, the ranking of Schmid factors in the $m'$ tables is only approximate. In some cases, such as the gold slip systems on either side of boundary 10 in Fig. 9 and Table 4, the most apparent slip traces (gold) are those with the lowest Schmid factors. Guery et al. (2016) similarly observed that even when local stress tensors extracted from a crystal plasticity simulation are used to compute Schmid factors, they are not well correlated with strain measurements using digital image correlation. Fig. 14 shows that use of the maximum $m'$ parameter for slip systems above a Schmid factor threshold of about 0.25 provides a meaningful way to predict the potential for slip transfer, but additional information such as the Schmid factor sum must also be considered. The geometrical relationship of $m'$ with misorientation is shown in Fig. 14, but it carries no information about activation of slip transfer, because the activation of slip systems in grains with a given misorientation with respect to the local stress state determines whether slip transmission will occur for combinations with a high $m'$. The geometry of slip transfer is convoluted with the driving force for slip in Fig. 14b, which suggests that despite the uncertainty of the local stress state and its effect on Schmid factors, there may be a meaningful threshold for slip transmission when considering both the geometry and the stress.

The appearance of aligned slip traces that are not compatible with each other such as the green and gold traces near grain boundary 25 in Fig. 8 indicates that the grain boundary was probably the source of dislocations, because the slip traces are uniform in topography both near and far from the boundary. For mutual nucleation of slip on incompatible planes from both sides of a grain boundary, sources at the surface in the boundary must be more easily activated than sources within the grain, suggesting that slip nucleated preferentially on the free surface, and progressed into the material below the surface. The evident ledge and the depression near the boundary in grain 2 (Fig. 8) indicates that the rate of dislocation production was greater in grain 2, as a greater strain would cause a greater reduction in the thickness of the grain. This is in contrast to grain boundary 4 (Fig. 7) that shows slip traces that fade as they get close to the boundary for low $m'$ conditions ($m' = 0.87$).

Fig. 15 shows a local average misorientation (LAM) map and a grain reference orientation deviation (GROD) map of the right side of the specimen. They show locations of high density of geometrically necessary dislocations (short-range orientation gradients) and the presence of long-range orientation gradients, respectively. The LAM map shows no systematic presence of high values at grain boundaries on the front surface. It is likely that much of the visible high value regions are associated with damage on the surface, such as the left half of grain 18, where there are many etch pit artefacts evident in Fig. 3a. Nevertheless, there is some accumulation of geometrical necessary dislocations near some boundaries on the back side. Assuming that the damage artefacts can be ignored, there is no systematic trend regarding local orientation gradients associated with either low or high angle boundaries.

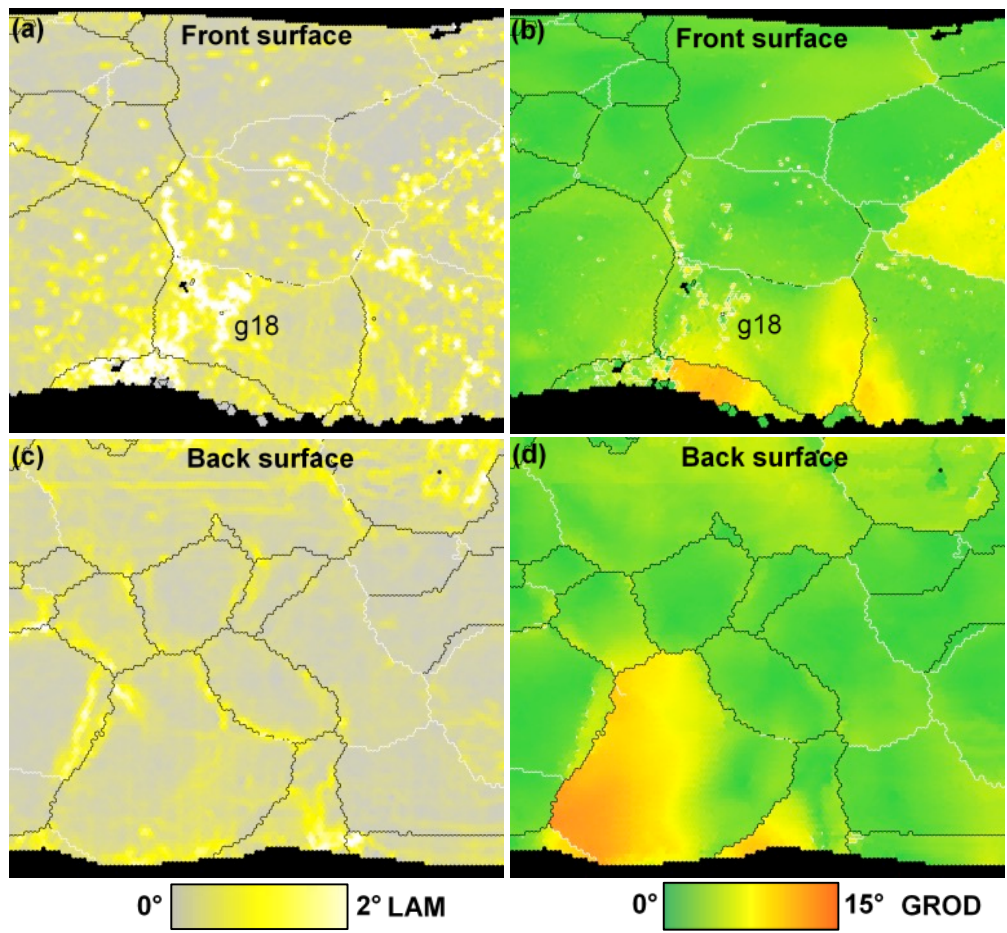

Fig. 15. (a,c) Local average misorientation (LAM, short range orientation gradients), and (b,d) grain reference orientation deviation (GROD, long-range orientation gradients) in the right side of the sample analysed. Results are shown for the front (a,b) and back (c,d) surfaces.

Similarly, the grain reference orientation deviation map also shows fairly uniform colors for each grain (the uniform differences in color are due to low angle boundaries), indicating a generally uniform orientation. The only exception is grain 18 where a neck was beginning to develop. Deformation to similar strains in BCC tantalum or hexagonal titanium show much more dramatic local orientation gradients, both along grain boundaries, and within the grains. This implies that deformation of near – cube orientated FCC grains is quite uniform despite the significant amount of slip traces.

There are some observations in this study that are not easy to explain, such as the white ridge in grain 2 in Fig. 8 or the lack of displacements in the shadowed slip traces by the gold slip traces in Fig. 10. Deviations from the ideal slip trace based upon a flat surface can seem rather large, and AFM or surface profilometry measurements would be useful to more clearly assess the magnitude of disagreement with ideal slip traces arising from heterogeneous surface topography and operation of invisible slip systems. It is also difficult to interpret regions that show extensive cross slip behaviour, such as the regions near grain boundary 42 in Fig. 3(a,b).

Taking these observations together, near-cube orientations do not consistently show slip transfer with $m' < 0.97$ (Fig. 14), due to many operational slip systems with large Schmid factors that enables strain self-accommodation. The observations of slip transfer in high angle grain boundaries are mostly correlated with the large grain 7, which is significantly different from a near-cube orientation. The size of the grain may also affect the operation of slip transfer in lower $m'$ conditions, due to the strength of the shear boundary conditions imposed by a large grain on smaller grains. The high hardening rates associated with the cube orientation due to activation of multiple slip systems leads to remarkable orientation stability, such that development of local orientation gradients is slower than in conditions where one slip system is prevalent. This is in contrast with other microstructures with more random textures and where one or two slip systems have much higher resolved shear stress or ease of operation, where slip transfer is more commonly observed (Hémery et al., 2018). Thus, finding the right value of $m'$ to use in a modeling setting appears to be situation dependent – if many slip systems are active in the neighborhood, then the $m'$ threshold is probably higher than when near-single slip conditions predominate.

The effect of deviation from a near cube orientation on convincing slip transfer is investigated in Fig. 16. In a ranked list of Schmid factors, an ideal cube orientation leads to eight equally highly stressed slip systems, but tilting away from the ideal cube orientation will increase the Schmid factor for some slip systems, and decrease others, which can be quantified by the slope in a plot such as in Fig. 16a for grain boundary 74, where slip transfer was observed with an $m'$ value of 0.94. In this boundary, one grain is more misoriented from the ideal cube orientation than the other, leading to a steeper slope for the more misoriented grain (each boundary has two slope values). Information from the $m'$ table is annotated and slip traces are provided in the inset to provide necessary context. The hypothesis that a greater slope would facilitate slip transfer at lower $m'$ values is assessed in Fig. 16b by plotting the gradient of Schmid factors with the observed $m'$ value, and in Fig. 16c by plotting the maximum Schmid factor against the misorientation; if slip transfer is observed at lower $m'$ values with greater slope or a higher Schmid factor, this would support the hypothesis. The 28 cases of convincing slip transfer and 24 cases of possible slip transfer are plotted with closed and open symbols, respectively, and there is no obvious way to assess the hypothesis. The boundaries that are at the extremes of the population (which are labelled) are associated with the large grain 7,. While the data used for this assessment are inconclusive, these data are from a limited subset of (mis)orientation space, so examination of such metrics with samples having a more random texture or a different loading direction is needed to further assess this hypothesis. Identification of a rule for an $m'$ threshold that is sensitive to slip system activity would provide useful guidance for installing slip transfer capability into computational models for polycrystal plasticity simulations.

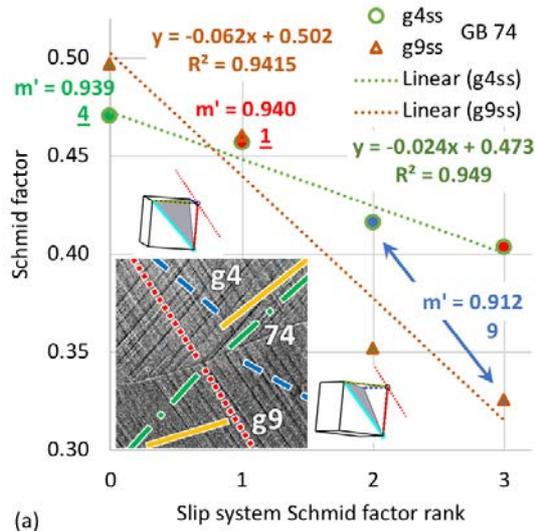

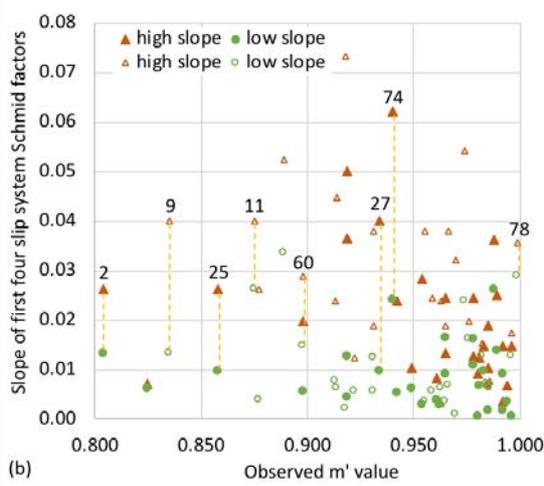

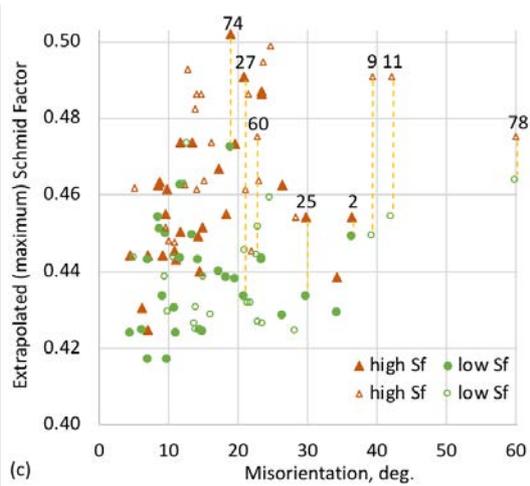

Fig. 16 (a) Deviation from the ideal cube orientation is related to the slope of Schmid factors with rank for grain boundary 74, where the slope is steeper for the more misoriented grain 9 (centers of symbols are colored according to the slip plane). (b) There is no obvious correlation between the slope and observed m' value, nor (c) between the maximum Schmid factor and the misorientation. There are two values for each boundary (brown triangles and green circles), and labeled instances connect these two values with gold dashed lines.

## 5. Conclusions

An assessment of slip transfer in near-cube FCC microstructures indicates that self-accommodation by slip is more easily activated than slip transfer when there are many slip systems available. Transfer across grain boundaries is rare, and only consistently evident when $m' > 0.97$, which corresponds to low-angle boundaries with <15º misorientation. What appears to be slip transfer in some instances could also be due to mutual nucleation of slip on both sides of the grain boundary from sources in the grain boundary on systems that have nearly parallel slip traces. The product of $m'$ times the sum of the Schmid factors of the two slip systems may provide a threshold for slip transfer, but this may only be meaningful for near-cube oriented grains. Non cube-oriented grains typically have a much wider range of Schmid factors on slip systems such that only one or two are highly active, so the threshold for slip transfer between such grains may have more complex $m'$ threshold criteria for slip transfer.

**Acknowledgements**

This investigation was supported by the European Research Council (ERC) under the European Union's Horizon 2020 research and innovation programme (Advanced Grant VIRMETAL, grant agreement No. 669141. T.R. Bieler acknowledges the support from the Talent Attraction program of the Comunidad de Madrid (reference 2016-T3/IND-1600) for his sabbatical in Madrid and R. Alizadeh also acknowledges the support from the Spanish Ministry of Science through the Juan de la Cierva program (FJCI-2016-29660). TRB also acknowledges support from the Department of Energy Office of Basic Science via grant DE-FG02-09ER46637.

**Appendix A. Grain orientation and selected grain boundary misorientation information**

All the EBSD data regarding the grain orientations and selected grain boundary misorientation information is presented in Tables A.1 and A.2. In Table A.1, the grain number is presented in first column, the average Euler angles corresponding to each grain are given in the next three columns (the Euler angle coordinate system has **X** down and **Y** to the right), and the **X** and **Y** position of the grain centroid are given in the last two columns (the image pixel (raster) coordinate system has **X** to the right and **Y** down). In Table A.2, the identity of selected grain boundaries is presented in column 1, the corresponding grains on the left (Grain L) and right (Grain R) side of the boundary are presented in column 2 as Grain L-Grain R numbers, the misorientation angle between two sides of the grain boundary is given in columns 3, the rotation axis in the crystal coordinate system (three integers) by which one grain can be rotated to have the same orientation as the other grain, is given in columns 4-6, and the calculated $m'$ values are listed in the last columns. Because "special boundaries" have tilt or twist rotations about low index rotation axes, it is apparent that there is no special pattern or similarity of rotation axes for any of the boundaries investigated except for boundary 78 on the left side, which is an annealing twin.

Table A.1. Average orientation, expressed in terms of Euler angles, and X-Y position (μm) of the grain centers presented in Figures 3 and 4

| Grain | $\varphi$ | $\phi$ | $\varphi_2$ | X | Y |
|---|---|---|---|---|---|
| **Left side** | | | | | |
| 1 | 105.4 | 156.7 | 282.2 | 183 | 80 |
| 2 | 122.3 | 170.2 | 299.4 | 559 | 126 |
| 3 | 25.8 | 171.5 | 207.8 | 1019 | 131 |
| 4 | 151.9 | 167.6 | 336.3 | 801 | 131 |
| 5 | 156.4 | 166.5 | 341.3 | 668 | 102 |
| 6 | 197.4 | 177.5 | 19.9 | 449 | 156 |
| 7 | 178.1 | 163.5 | 19.3 | 173 | 480 |
| 8 | 325.7 | 161.8 | 49.2 | 272 | 212 |
| 9 | 181.7 | 151.7 | 8.1 | 915 | 228 |
| 10 | 157.9 | 174.6 | 338.2 | 584 | 377 |
| 11 | 57.3 | 173.6 | 237.6 | 349 | 343 |
| 12 | 263.4 | 169.6 | 91.6 | 868 | 399 |
| 13 | 278.7 | 176.0 | 13.6 | 1013 | 374 |
| 14 | 132.3 | 142.7 | 280.4 | 819 | 496 |
| 15 | 219.5 | 169.5 | 33.5 | 994 | 521 |
| 16 | 8.3 | 164.4 | 198.8 | 708 | 559 |
| 17 | 312.6 | 169.8 | 38.8 | 1100 | 501 |
| 18 | 274.1 | 165.9 | 96.3 | 762 | 702 |
| 19 | 359.0 | 140.9 | 340.3 | 1085 | 650 |
| 20 | 331.7 | 168.9 | 69.1 | 348 | 729 |
| 21 | 188.4 | 156.2 | 31.8 | 40 | 776 |
| 22 | 244.0 | 168.9 | 63.3 | 1075 | 724 |
| 23 | 333.7 | 161.5 | 63.1 | 1075 | 766 |
| 24 | 49.2 | 133.3 | 266.0 | 209 | 795 |
| **Right side** | | | | | |
| 1 | 4.4 | 8.3 | 354.4 | 82 | 26 |
| 2 | 349.2 | 11.4 | 5.0 | 237 | 39 |
| 3 | 149.2 | 14.9 | 209.5 | 472 | 60 |
| 4 | 144.7 | 15.9 | 199.2 | 784 | 11 |
| 5 | 107.6 | 18.5 | 247.8 | 801 | 44 |
| 6 | 257.0 | 4.0 | 94.6 | 157 | 55 |
| 7 | 80.8 | 7.4 | 281.5 | 692 | 149 |
| 8 | 120.3 | 17.5 | 239.1 | 7 | 64 |
| 9 | 191.1 | 15.6 | 174.3 | 76 | 112 |
| 10 | 261.4 | 10.7 | 101.8 | 214 | 157 |
| 11 | 120.5 | 1.8 | 242.6 | 483 | 147 |
| 12 | 335.1 | 8.1 | 26.7 | 45 | 208 |
| 14 | 116.8 | 11.9 | 242.1 | 442 | 283 |
| 15 | 33.7 | 1.5 | 325.7 | 733 | 439 |
| 16 | 176.5 | 38.9 | 200.3 | 127 | 395 |
| 17 | 330.6 | 9.4 | 32.6 | 657 | 271 |
| 18 | 83.1 | 18.3 | 279.4 | 453 | 480 |
| 19 | 118.4 | 8.2 | 264.1 | 300 | 577 |
| 21 | 160.4 | 15.0 | 219.0 | 585 | 635 |

Table A.2. Misorientations and rotation axes of selected boundaries in Figures 3 and 4, together with the value of $m'$ associated with observed slip transfer (gray values indicate less certainty)

| Boundary | Grain pair | Misorientation | Rotation axis | | | $m'$ obs. |
|---|---|---|---|---|---|---|
| **Left side** | | | | | | |
| 46 | 6 - 10 | 4.5 | -6 | -6 | -5 | 0.996 |
| 52 | 2 - 10 | 7.1 | 0 | -9 | -5 | 0.992 |
| 4 | 21 - 7 | 8.6 | -7 | 9 | 4 | 0.988 |
| 39 | 11 - 10 | 9.1 | -25 | 10 | 1 | 0.983 |
| 42 | 6 - 2 | 11.0 | 3 | 10 | 6 | 0.98 |
| 35 | 1 - 2 | 14.2 | -1 | -30 | -2 | 0.965 |
| 105 | 15 - 13 | 14.4 | 2 | 7 | 9 | |
| 91 | 18 - 15 | 14.8 | -15 | -13 | -16 | 0.961 |
| 89 | 9 - 13 | 15.4 | -4 | -10 | -7 | 0.962 |
| 47 | 10 - 18 | 17.4 | -5 | -12 | -2 | |
| 74 | 4 - 9 | 18.9 | 10 | -5 | -2 | 0.94 |
| 27 | 8 - 11 | 20.9 | -4 | -6 | 3 | 0.934 |
| 33 | 8 - 6 | 21.6 | -6 | 10 | -5 | |
| 60 | 10 - 16 | 22.7 | 10 | -1 | 5 | 0.898 |
| 57 | 16 - 18 | 23.9 | -5 | 7 | -4 | |
| 37 | 7 - 10 | 24.2 | -9 | 3 | 17 | |
| 34 | 1 - 6 | 24.3 | 2 | 12 | 3 | |
| 73 | 10 - 9 | 24.4 | 11 | -2 | -3 | |
| 43 | 8 - 10 | 24.4 | -2 | 3 | -1 | |
| 72 | 2 - 9 | 27.2 | 12 | -5 | -6 | |
| 45 | 7 - 18 | 28.3 | 11 | 6 | -9 | 0.877 |
| 89 | 9 - 13 | 29.1 | 0 | 30 | -1 | |
| 25 | 7 - 11 | 29.8 | -7 | 3 | 8 | 0.858 |
| 15 | 7 - 20 | 29.9 | 1 | 21 | -10 | |
| 2 | 7 - 1 | 36.5 | -2 | 7 | 8 | 0.804 |
| 9 | 1 - 8 | 39.3 | -7 | -13 | 2 | 0.835 |
| 11 | 7 - 8 | 42.2 | -1 | 5 | -4 | 0.875 |
| 122 | 18 - 19 | 42.9 | -16 | 3 | 6 | |
| 7 | 7 - 24 | 43.3 | 7 | 11 | -3 | |
| 77 | 10 - 14 | 45.8 | 5 | 12 | 13 | |
| 78 | 16 - 14 | 60.0 | -10 | -9 | -10 | 0.999 |
| **Right side** | | | | | | |
| | Within 15 | 6 | 26 | 8 | 3 | |
| 38 | 11 - 7 | 6.2 | 9 | 22 | -3 | 0.994 |
| 50 | 7 - 15 | 7.1 | 0 | -7 | -3 | 0.992 |
| 48 | 17 - 15 | 9.7 | 16 | -13 | 9 | 0.985 |
| 33 | 14 - 11 | 10.9 | 6 | -13 | 6 | 0.981 |
| 27 | 14 - 18 | 11.5 | 18 | 18 | 11 | |
| 46 | 17 - 7 | 13.9 | 9 | -17 | 2 | 0.965 |
| 31 | 3 - 11 | 14.1 | -13 | 7 | -5 | 0.966 |
| 52 | 7 - 5 | 14.4 | 14 | -21 | 15 | 0.962 |
| 39 | 3 - 7 | 14.6 | 28 | -1 | 9 | 0.955 |
| 10 | 9 - 10 | 16.0 | 23 | -15 | -6 | |
| 54 | 4 - 5 | 17.1 | -10 | -17 | -23 | 0.954 |
| 42 | 18 - 15 | 17.6 | -2 | -25 | -4 | |
| 17 | 10 - 2 | 18.2 | 10 | 7 | -8 | |
| 44 | 14 - 17 | 21.0 | -11 | 13 | -4 | 0.931 |
| 24 | 10 - 3 | 21.6 | 10 | -17 | 3 | 0.931 |
| 19 | 10 - 14 | 21.8 | -4 | 25 | -4 | 0.922 |

| 22 | 19 - 18 | 22.9 | 10  | 11  | -23 | 0.918 |
| 3  | 12 - 9  | 23.0 | 27  | -2  | -5  |       |
| 43 | 21 - 15 | 25.4 | 17  | -8  | -23 |       |
| 25 | 2 - 3   | 26.2 | -6  | 2   | 1   |       |
| 8  | 16 - 19 | 36.0 | -9  | 2   | -2  |       |
| 18 | 16 - 14 | 37.0 | -22 | -2  | 9   |       |
| 11 | 16 - 10 | 42.7 | 20  | -11 | -10 |       |
| 20 | 16 - 18 | 44.0 | -11 | -2  | 2   |       |
| 4  | 12 - 16 | 48.7 | 27  | -1  | 12  |       |

**References**

Abu Al-Rub, R.K., Voyiadjis, G.Z., 2006. A physically based gradient plasticity theory. Int. J. Plast. 22, 654–684.

Abuzaid, W.Z., Sehitoglu, H., Lambros, J., 2016. Localisation of plastic strain at the microstructural level in Hastelloy X subjected to monotonic, fatigue, and creep loading: the role of grain boundaries and slip transmission. Mater. High Temp. 33, 384–400.

Bayerschen, E., McBride, A.T., Reddy, B.D., Böhlke, T., 2016. Review on slip transmission criteria in experiments and crystal plasticity models. J. Mater. Sci. 51, 2243–2258.

Bieler, T.R., Eisenlohr, P., Zhang, C., Phukan, H.J., Crimp, M.A., 2014. Grain boundaries and interfaces in slip transfer. Curr. Opin. Solid St. M. 18, 212–226.

Boehlert, C., Chen, Z., Gutirrez-Urrutia, I., Llorca, J., Pérez-Prado, M., 2012. In situ analysis of the tensile and tensile-creep deformation mechanisms in rolled AZ31. Acta Mater. 60, 1889–1904.

Bond, D. M., Zikry, M.A., 2017. A predictive framework for dislocation-density pile-ups in crystalline systems with coincident site lattice and random grain boundaries. J. Eng. Mater. Technol. 139(2), 021023.

Busso, E.P., Meissonnier, F.T., O'Dowd, N.P., 2000. Gradient-dependent deformation of two-phase single crystals. J. Mech. Phys. Solids. 48, 2333–2361.

Chandra, S., Samal, M.K., Chavan, V.M., Patel, R.J., 2015. Atomistic simulations of interaction of edge dislocation with twist grain boundaries in Al-effect of temperature and boundary misorientation. Mater. Sci. Eng. A. 646, 25–32.

Chandra, S., Samal, M.K., Chavan, V.M., Patel, R.J., 2016. An atomistic study of resistance offered by twist grain boundaries to incoming edge dislocation in FCC metals. Mater. Lett. 180, 11–14.

Chen, Z., Daly, S.H., 2017. Active slip system identification in polycrystalline metals by digital image correlation (DIC). Exp. Mech. 57, 115–127.

Chen, B., Jiang, J., Dunne, F.P.E., 2018. Is stored energy density the primary meso-scale mechanistic driver for fatigue crack nucleation?. Int. J. Plast. 101, 213–229.

Cruzado, A., Lucarini, S., LLorca, J., Segurado, J., 2018. Microstructure-based fatigue life model of metallic alloys with bilinear Coffin-Manson behavior. Int. J. Fatigue. 107, 40–48.

Delaire, F., Raphanel, J.L., Rey, C., 2000. Plastic heterogeneities of a copper multicrystal deformed in uniaxial tension: experimental study and finite element simulations. Acta Mater. 48, 1075–1087.

Dewald, M.P., Curtin, W.A., 2007. Multiscale modelling of dislocation/grain boundary interactions. II. Screw dislocations impinging on tilt boundaries in Al. Philos. Mag. 87, 4615–4641.

Ding, R., Gong, J., Wilkinson, A.J., Jones, I.P., 2016. A study of dislocation transmission through a grain boundary in hcp Ti–6Al using micro-cantilevers. Acta Mater. 103, 416–423.

Eftink, B.P., Li, A., Szlufarska, I., Mara, N.A., Robertson, I.M., 2017. Deformation response of AgCu interfaces investigated by in situ and ex situ TEM straining and MD simulations. Acta Mater. 138, 212–223.

Guery, A., Hild, F., Latourte, F., Roux, S., 2016. Slip activities in polycrystals determined by coupling DIC measurements with crystal plasticity calculations. Int. J. Plast. 81, 249–266.

Gurtin, M.E., 2002. A gradient theory of single-crystal viscoplasticity that accounts for geometrically necessary dislocations. J. Mech. Phys. Solids. 50(1), 5–32.

Gurtin, M.E., 2008. A theory of grain boundaries that accounts automatically for grain misorientation and grain-boundary orientation. J. Mech. Phys. Solids. 56(2), 640–662.

Haouala, S., Segurado, J., LLorca, J., 2018. An analysis of the influence of grain size on the strength of FCC polycrystals by means of computational homogenization. Acta Mater. 148, 72–85.

Hémery, S., Nizou, P., Villechaise, P., 2018. In situ SEM investigation of slip transfer in Ti-6Al-4V: Effect of applied stress. Mater. Sci. Eng. A. 709, 277–284.

Kacher, J., Eftink, B.P., Cui, B., Robertson, I.M., 2014. Dislocation interactions with grain boundaries. Curr. Opin. Solid St. M..18, 227–243.

Koning, M.D., Miller, R., Bulatov, V., Abraham, F.F., 2002. Modelling grain-boundary resistance in intergranular dislocation slip transmission. Philos. Mag. A. 82, 2511–2527.

Lebensohn, R.A., Tome, C.N., 1993. A self-consistent anisotropic approach for the simulation of plastic deformation and texture development of polycrystals: application to zirconium alloys, Acta Metall. Mater. 41(9), 2611–2624.

Lee, T.C., Robertson, I.M., Birnbaum, H.K., 1989. Prediction of slip transfer mechanisms across grain boundaries. Scripta Metall. 23, 799–803.

Lim, H., Lee, M.G., Kim, J.H., Adams, B.L., Wagoner, R.H., 2011. Simulation of polycrystal deformation with grain and grain boundary effects. Int. J. Plast. 27, 1328–1354.

Lim, H., Carroll, J.D., Battaile, C.C., Boyce, B.L., Weinberger, C.R., 2015. Quantitative comparison between experimental measurements and CP-FEM predictions of plastic deformation in tantalum oligocrystal. Int. J. Mech. Sci. 92, 98–108.

Liu, B., Eisenlohr, P., Roters, F., Raabe, D., 2012. Simulation of dislocation penetration through a general low-angle grain boundary. Acta Mater. 60, 5380–5390.

Livingston, J.D., Chalmers, B., 1957. Multiple slip in bicrystal deformation. Acta Metall. 5(6), 322–327.

Luster, J., Morris, M., 1995. Compatibility of deformation in two-phase Ti-Al alloys: dependence on microstructure and orientation relationships. Metall. Mater. Trans. A 26(7), 1745–1756.

Ma, A., Roters, F., Raabe, D., 2006. On the consideration of interactions between dislocations and grain boundaries in crystal plasticty finite element modeling – Theory, experiments, and simulations. Acta Mater. 54, 2181–2194.

Malyar, N., Micha, J., Dehm, G., Kirchlechner, C., 2017. Size effect in bicrystalline micropillars with a penetrable high angle grain boundary. Acta Mater. 129, 312–320.

Muñoz-Moreno, R., Boehlert, C.J., Pérez-Prado, M.T., Ruiz-Navas, E.M., LLorca, J., 2013. Effect of stress level on the high temperature deformation and fracture mechanisms of Ti-45Al-2Nb-2Mn-0.8v.%TiB2: an in situ experimental study. Metall. Mater. Trans. A, 44, 1887–1896.

Musinski, W.D., McDowell, D.L., 2016. Simulating the effect of grain boundaries on microstructurally small fatigue crack growth from a focused ion beam notch through a three-dimensional array of grains. Acta Mater. 112, 20–39.

Roters, F., Eisenlohr, P., Hantcherli, L., Tjahjanto, D.D., Bieler, T.R., Raabe, D., 2010. Overview of constitutive laws, kinematics, homogenization, and multiscale methods in crystal plasticity finite element modeling: theory, experiments, applications. Acta Mater. 58(4), 1152–1211.

Rubio, R. A., Haouala, S., LLorca, J., 2019, . Grain boundary strengthening of FCC polycrystals. Journal of Materials Research, 34, In press.

Sangid, M.D., Maier, H.J., Sehitoglu, H., 2011. The role of grain boundaries on fatigue crack, initiation – An energy approach. Int. J. Plast. 27, 801–821.

Segurado, J., Lebensohn, R.A., LLorca, J., 2018. Computational homogenization of polycrystals. Adv. Appl. Mech. 51, 1-114.

Spearot, D.E., Sangid, M.D., Insights on slip transmission at grain boundaries from atomistic simulations. Curr. Opin. Solid St. M. 18, 188–195.

Stinville, J.C., Vanderesse, N., Bridier, F., Bocher, P., Pollock, T.M., 2015. High resolution mapping of strain localization near twin boundaries in a nickel-based superalloy. Acta Mater. 98, 29–42.

Stinville, J.C., Lenthe, W.C., Echlin, M.P., Callahan, P.G., Texier, D., Pollock, T.M., 2017. Microstructural statistics for fatigue crack initiation in polycrystalline nickel-base superalloys. Int. J. Fracture, 208, 221–240.

Su, Y., Zambaldi, C., Mercier, D., Eisenlohr, P., Bieler, T.R., Crimp, M.A., 2016. Quantifying deformation processes near grain boundaries in a titanium using nanoindentation and crystal plasticity modeling. Int. J. Plast. 86, 170–186.

Tschopp, M.A., Spearot, D.E., McDowell, D.L., 2008. Influence of grain boundary structure on dislocation nucleation in FCC metals. Dislocations in Solids. Edited by J. P. Hirth, Elsevier B.V.

Tsuru, T., Shibutani, Y., Hirouchi, T., 2016. A predictive model for transferability of plastic deformation through grain boundaries. AIP Adv. 6, 015004.

van Beers, P.R.M., McShane, G.J., Kouznetsova, V.G., Geers, M.G.D., 2013. Grain boundary interface mechanics in strain gradient crystal plasticity. J. Mech. Phys. Solids. 61, 2659–2679.

Wan, V., MacLachlan, D., Dunne, F., 2016. Integrated experiment and modelling of microstructurally-sensitive crack growth. Int. J. Fatigue. 91, 110–123.

Wang, L., Yang, Y., Eisenlohr, P., Bieler, T.R., Crimp, M.A., Mason, D.E., 2010. Twin nucleation by slip transfer across grain boundaries in commercial purity titanium. Metall. Mat. Trans A. 41, 421–30.

Xiao, X., Terentyev, D., Chen, Q., Yu, L., Chen, L., Bakaev, A., Duan, H., 2017. The depth dependent hardness of bicrystals with dislocation transmission through grain boundaries: A theoretical model. Int. J. Plast. 90, 212–230.

Xu, S., Xiong, L., Chen, Y., McDowell, D.L., 2016. Sequential slip transfer of mixed-character dislocations across Σ3 coherent twin boundary in FCC metals: a concurrent atomistic-continuum study. npj Comp. Mater. 2, 15016.

Yang, Y., Wang, L., Bieler, T.R., Eisenlohr, P., Crimp, M.A., 2011. Quantitative atomic force microscopy characterization and crystal plasticity finite element modeling of heterogeneous deformation in commercial purity titanium, Metall. Mater. Trans A, 42A(3), 636–644.

Zhang, L., Lu, Ch., Tieu, K., 2016. A review on atomistic simulation of grain boundary behaviors in face-centered cubic metals. Comp. Mater. Sci. 118, 180–191.